\documentclass[preprint, amsmath,amssymb, aps]
{revtex4-2}

\usepackage{graphicx}
\usepackage{dcolumn}
\usepackage{bm}

\usepackage{xcolor} 
\usepackage{subcaption} 
\usepackage{algorithm2e}
\usepackage{mathrsfs}

\begin{document}

\preprint{APS/123-QED}

\title{Simulating squirmers with smoothed particle dynamics}

\author{Xinwei Cai}
\affiliation{State Key Laboratory of Fluid Power and Mechatronic Systems, Department of Engineering Mechanics, Zhejiang University, Hangzhou 310027, PR China\\}%
\author{Kuiliang Wang}
\affiliation{State Key Laboratory of Fluid Power and Mechatronic Systems, Department of Engineering Mechanics, Zhejiang University, Hangzhou 310027, PR China\\}%
\author{Gaojin Li}
\affiliation{State Key Laboratory of Ocean Engineering, School of Ocean and Civil Engineering, Shanghai Jiaotong University, Shanghai 200240, PR China\\}

\author{Xin Bian}%
 \email{bianx@zju.edu.cn}
\affiliation{State Key Laboratory of Fluid Power and Mechatronic Systems, Department of Engineering Mechanics, Zhejiang University, Hangzhou 310027, PR China}%




\date{\today}

\begin{abstract}

Microswimmers play an important role in shaping the world around us. The squirmer is a simple model for microswimmer whose cilia oscillations on its spherical surface induce an effective slip velocity to propel itself. The rapid development of computational fluid dynamics methods has markedly enhanced our capacity to study the behavior of squirmers in aqueous environments. Nevertheless, a unified methodology that can fully address the complexity of fluid-solid coupling at multiple scales and interface tracking for multiphase flows remains elusive, posing an outstanding challenge to the field.  
To this end, we investigate the potential of the smoothed particle dynamics~(SPD) method as an alternative approach for simulating squirmers. The Lagrangian nature of the method allows it to effectively address the aforementioned difficulty. By introducing a novel treatment of the boundary condition and assigning appropriate slip velocities to the boundary particles,
the SPD-squirmer model is able to accurately represent a range of microswimmer types including pushers, neutral swimmers, and pullers. We systematically validate the steady-state velocity of the squirmer, the resulting flow field, its hydrodynamic interactions with the surrounding environment, and the mutual collision of two squirmers. In the presence of Brownian motion, the model is also able to correctly calculate the velocity and angular velocity autocorrelation functions at the mesoscale. Finally, we simulate a squirmer within a multiphase flow by considering a droplet that encloses a squirmer and imposing a surface tension between the two flow phases. We find that the squirmer within the droplet exhibits different motion types. 
Since the proposed method is applicable to a wide range of complex scenarios, it has implications for a number of areas, including the design and application of micro/nano artificial swimmers, flow manipulation in microfluidic chips, and drug delivery in the biomedical field.

\end{abstract}

\maketitle


\section{\label{sec:level1}Introduction}

In examining the microscopic realm of life, it is crucial to acknowledge the minuscule yet profoundly influential entities known as microorganisms. Although their existence is largely invisible to the naked eye, they are found in astonishing diversity and abundance throughout the world. From the depths of marine ecosystems to the inner workings of the human gut, these microorganisms utilize their unique motility to navigate and reproduce in their environments.
They also play a critical role in energy flow, matter cycling and the complex interplay between health and disease~\cite{ingraham2003introduction,Mo2023_Li_Bian_FrontiersInPhysics,ishikawa2024_AnnuRevFluidMech}. The movement of microorganisms in aqueous environments is dominated by viscous forces rather than inertial force due to their small size~\cite{purcell2014life}, which has led to the evolution of specialized locomotive structures such as flagella and cilia~\cite{khan2018_scholey}. These organelles, which are characterized by a 9+2 microtubule arrangement, enable microorganisms to navigate and sense their surroundings effectively~\cite{fisch2011_Dupuis_BioCell}.
Flagella are typically long and sparsely distributed, such as in Escherichia coli, Sperm cells and Chlamydomonas~\cite{elgeti2015_et_al,drescher2010_et_al_PhysRevLett}. 
Cilia, on the other hand, are relatively short and densely packed, often covering the entire surface of the microorganism. For example, Opalina, Paramecium, and certain green algae move forward by the coordinated beating of their cilia~\cite{tamm1970_horridge,hausmann2010electron,kirk1998volvox}. In these organisms, each cilium performs an effective stroke followed by a recovery, collectively forming a metachronal wave that generates propulsive force~\cite{brennen1974_JFluidMech,michelin2010_lauga_PhysFluid,brumley2012_PhysRevLett}. The morphologies of motile microorganisms are diverse, ranging from a few micrometers to several millimeters in size, and their motility behavior can also change in response to environmental variations. Mathematical modeling of these microorganisms poses a formidable challenge, requiring the development of simplified models for study.

In the early 1950s, mathematical modeling of low-Reynolds-number swimmers began to emerge. Taylor demonstrated through the propulsion mechanism of a long sperm tail that an infinitely deformable sheet immersed in a viscous fluid can propel itself to the left by generating small-amplitude transverse waves that propagate to the right, with the propulsion velocity proportional to the square of the wave amplitude~\cite{taylor1951analysis}. Subsequently, Lighthill introduced an idealized model for a finite body: the spherical squirmer~\cite{lighthill1952},
where he envisioned a sphere with a deformable surface and hypothesized small-amplitude radial and tangential, axisymmetric, and periodic motions of the surface elements. Blake further developed the squirmer model and applied it to the study of ciliate motility, considering the deformable and stretchable surface of the sphere as the envelope of the cilia tips during their beating motion~\cite{blake1971_JFluidMech}.
The squirmer model allows an effective slip velocity directly on the surface of the sphere to represent the flow induced by the cilia, 
which has been widely used to analyze various behaviors and properties of microswimmers. 
For example, Magar et al.~\cite{magar2003_et_al_QJMechAppMath} used this model to analyze the nutrient uptake characteristics of solitary swimmers. Over the past decade, researchers have employed the model for a plethora of studies, including the analysis of swimming efficiency~\cite{michelin2010_lauga_PhysFluid,michelin2013_lauga_JFluidMech}, the motion characteristics of squirmers in confined geometries, such as near walls~\cite{li2014_PhysRevE} and free surfaces~\cite{ishimoto2013_gaffney_PhysRevE}, as well as hydrodynamic interactions between two swimmers~\cite{ishikawa2006_et_al_JFluidMech, li2016_PhysRevE}. In addition, the motion of swimmers in complex fluids has been studied, including their motion in density gradients~\cite{shaik2021_ardekani_PhysFluid}, viscosity gradients~\cite{datt2019_elfring_PhysRevLett}, and non-Newtonian fluids~\cite{zhu2012_et_al_PhysFluid, ouyang2023_et_al_JFluidMech}. Furthermore, to gain a deeper understanding of the collective dynamics of microswimmers, the group behaviors of numerous squirmers have also been studied~\cite{ishikawa2008_pedley_PhysRevLett, zottl2014_stark_PhysRevLett,kyoya2015_et_al_PhysRevE}.

The rapid development of computational fluid dynamics methods has greatly facilitated the simulation of microswimmers. Popular methods include lattice Boltzmann method~\cite{kvs2021_et_al_PhysRevE,nie2024_et_al_PhysRevE}, dissipative particle dynamics~\cite{bailey2024_et_al_Nanoscale,overberg2024_gompper_SoftMatter}, finite volume method~\cite{li2014_PhysRevE,li2016_PhysRevE}, and multi-particle collision dynamics~\cite{downton2009_stark_JPhysCondMatter,gotze2010_gompper_PhysRevE,zottl2018_stark_EurPhysE}. 
Each of these approaches has distinct advantages and has individually achieved considerable success. Despite these collective advances, a unified methodology that effectively addresses the complexity of fluid-solid coupling at multiple scales and interface tracking for multiphase flows remains elusive. This gap presents a formidable challenge that requires innovative solutions. In direct response to this challenge, we apply the smoothed particle dynamics~(SPD) method to model microswimmers. Capitalizing on the Lagrangian framework inherent in SPD, this approach presents a unified and potent strategy to overcome the previously described complexity.

SPD refers to either smoothed particle hydrodynamics~(SPH) or smoothed dissipative particle dynamics~(SDPD), which are utilized to solve either macroscopic or mesoscopic flow problems, respectively. The Lagrangian characteristic of SPD is advantageous for handling complex interfaces, including fluid-solid coupling with moving boundaries and interface tracking. Originally developed to simulate phenomena in astrophysics, SPH has since been extensively applied to a range of flow problems~\cite{liu2010_Liu_ACompuME,monaghan2012smoothed,zhang2022_et_al_JHydro}. In the past few decades, SPH has been effectively employed to tackle complex flow problems, particularly those involving multiphase flow~\cite{hu2006_adams_JCompPhys, wang2023_et_al_Frontiers,zheng2024_et_al_AQUA} and particle suspensions~\cite{bian2012_et_al_PhysFluid, Vazquez_et_al2016_CompParticleMechnics,Cai2024_Li_Bian}. SDPD, proposed by Espa{\~n}ol and Revenga~\cite{espanol2003smoothed}, introduces stochastic forces into SPH within the GENERIC framework of thermodynamics~\cite{Grmela1997}, making it an effective solver for the Landau-Lifshitz-Navier-Stokes equations~\cite{landau1959,bian2015fluctuating,bian2016pre,ellero2018everything}.
It has been widely applied to study the physics of various mesoscopic flows~\cite{hu2006_adams_JCompPhys,litvinov2008smoothed,Vazquez-Quesada2009a,bian2012_et_al_PhysFluid,Ye2017}. Building on our previous work on the slip boundary condition~\cite{Cai2023_Li_Bian_JCompPhys}, we model the dynamics of squirmers for a variety of fluid dynamics problems ranging from the macroscopic to the mesoscopic scale, and from single phase to multiple phases.

The structure of the paper is as follows: in section~\ref{section_method} we present the continuous and discrete forms of the fluid dynamics equations and provide a detailed exposition on how to construct the squirmer model within the SPD method. In section~\ref{section_results} we demonstrate the accuracy and versatility of the model through several flow problems. Finally in section~\ref{section_conclusion} we summarize this work.

\section{The method}\label{section_method}

\subsection{The squirmer model}

The squirmer is a simple model for microswimmer, originally used to model the locomotion and hydrodynamics of swimming ciliated microorganisms. 
Lighthill~\cite{lighthill1952} first introduced the Squirmer model, which describes the locomotion of a deformable body swimming forward through small oscillations at low Reynolds numbers. Subsequently, Blake~\cite{blake1971_JFluidMech} refined this model by approximating the flow induced by the periodic oscillation of cilia enveloping the sphere's surface with an effective slip velocity. In this theoretical setup, the trajectory of a particle on the surface of a squirmer is defined by the tips of the cilia. The flow adjacent to the surface can be characterized as an effective slip velocity $\mathbf{v}_s$ directly on the surface of the sphere. The slip velocity can be written as an infinite series of eigenfunctions of the Stokes equation describing arbitrary time-dependent squirming velocities~\cite{ishikawa2024_AnnuRevFluidMech}. For axisymmetric flows, the surface velocity field with the polar angle $\theta$ and time $t$ is simplified in spherical coordinates as
\begin{equation}
    \mathbf{v}_s(\theta, t) = v_{\theta}(\theta, t)\mathbf{e}_{\theta} + v_r(\theta, t)\mathbf{e}_r.
\end{equation}
Here, the polar angle $\theta$ is defined by the direction of the head $\mathbf{e}$ and the radial vector $\mathbf{r}_s$ from the center of the sphere to the point on the surface. $v_{\theta}$ and $v_r$ correspond to the velocity components in the directions tangential and radial to the surface, respectively.

The original squirmer model did not constrain the nature of the surface velocity. However, for the sake of simplicity, subsequent research often assumes a time-independent, axisymmetric and tangential surface velocity. For an accurate representation of the flow field, it is sufficient to consider only the first two modes of the tangential velocity, $B_1$ and $B_2$~\cite{blake1971_JFluidMech}. In this simplified framework, the surface velocity, which manifests exclusively in the tangential direction on the sphere, can be described by 
\begin{equation}
    \mathbf{v}_s(\theta)=v_{\theta}(\theta)\mathbf{e}_{\theta} = B_1(sin\theta + \beta sin\theta cos\theta)\mathbf{e}_{\theta}.
    \label{vs_theta}
\end{equation}
where $\beta$ represents the squirmer parameter, defined as
\begin{equation}
    \beta = \frac{B_2}{|B_1|},
\end{equation}
which quantifies the leading-order flow field: pushers ($\beta < 0$), pullers ($\beta > 0$), and neutral swimmers ($\beta = 0$). Eq.~(\ref{vs_theta}) has another form:
\begin{equation}
    \mathbf{v}_s(\mathbf{r}_s, \mathbf{e}) = B_1 \left[1+\beta\left(\mathbf{e}\cdot \frac{\mathbf{r}_s}{R}\right) \right]\left[\left(\mathbf{e}\cdot \frac{\mathbf{r}_s}{R}\right)\frac{\mathbf{r}_s}{R}-\mathbf{e}\right],
    \label{surface_velocity}
\end{equation}
where $R = |\mathbf{r}|$ is the radius of the sphere.

For this model the velocity of the swimmer $U_0$ in bulk is solely determined by the first mode $B_1$ as~\cite{blake1971_JFluidMech}
\begin{eqnarray}
    U_0 = \frac{2}{3}|B_1|.
\end{eqnarray}

Considering the translation and rotation of the squirmer, the absolute velocity on the surface of the sphere is given by
\begin{equation}
    \mathbf{v} = \mathbf{U}_0 + \mathbf{\Omega}_0 \times \mathbf{r}_s + \mathbf{v}_s(\mathbf{r_s},\mathbf{e}),
    \label{absolute_velocity}
\end{equation}
where $\mathbf{U}_0$, $\mathbf{\Omega}_0$ are the translation and angular velocity of the sphere, respectively.

The kinematics of the squirmer is governed by the principles of rigid body dynamics. Its translational and rotational movements are described by the classical Newton's second law and Euler's equations of motion, as follows:
\begin{eqnarray}
\mathbf{F} &=&  M\dot{\mathbf{U}}_0\\
\mathbf{T} &=& \mathbf{I}\cdot\dot{\mathbf{\Omega}}_0+\mathbf{\Omega}_0\times[\mathbf{I}\cdot\mathbf{\Omega}_0]
\end{eqnarray}
Here, $\mathbf{F}$ and $\mathbf{T}$ denote the net force and torque acting on the squirmer, respectively. $M$ represents the total mass of the squirmer, while $\mathbf{I}$ signifies the inertia tensor that encapsulates the distribution of mass relative to the squirmer's center of mass.

\subsection{Lagrangian hydrodynamic equations}

The equations of isothermal Newtonian fluid in a Lagrangian description are
\begin{eqnarray}
   \frac{{\mathrm{d}}\rho}{{\mathrm{d}} t} & =&  -\rho \nabla \cdot \mathbf{v},\label{continuity_eq} \\
    \rho \frac{{\mathrm{d}}\mathbf{v}}{{\mathrm{d}}t} &=& -\nabla p + \nabla \cdot [\eta(\nabla \mathbf{v}+\nabla \mathbf{v}^T)] - \frac{2}{3}\nabla(\eta\nabla\cdot \mathbf{v}) + \rho\mathbf{g} + \mathbf{F}^s, \label{ns_eq}
\end{eqnarray}
where $\rho$, $\mathbf{v}$, $p$, $\eta$, and $\mathbf{g}$ are material density, velocity, pressure, dynamic viscosity and body force per unit mass, respectively. An equation of state (EOS) relating the pressure to the density is necessary to provide a closure for
a weakly compressible description, and it can be expressed as:
\begin{equation}
p=c_{s}^2 \rho_{0}\left[\left(\frac{\rho}{\rho_{0}}\right)-1\right]+\chi \label{eoc_eq},
\end{equation}
where $\rho_0$ is the equilibrium density. An artificial sound speed $c_s$ is chosen based on a scale analysis~\cite{Morris1997_et_al_JCompPhys} such that the pressure field reacts strongly to small deviations in the density, and therefore a quasi-incompressibility is fulfilled. 
In this case, the third term on the rhs. of Eq.~(\ref{ns_eq}) may be negligible.
Here, $\chi$ is a positive constant introduced to enforce the non-negativity of pressure on discrete SPD particles.

$\mathbf{F}^s$ in Eq.~(\ref{ns_eq}) represents a surface force which acts at the surface between two immiscible fluid phases as follows
\begin{eqnarray}
    \mathbf{F}^s=\alpha \kappa \mathbf{n},
\end{eqnarray}
where $\alpha$, $\kappa$, $\mathbf{n}$ are the surface tension coefficient, the curvature of the interface and the unit normal vector at the interface, respectively. The normal vector can be obtained using $\mathbf{n} = \frac{\nabla C}{|\nabla C|}$, where $C$ is a colour function that has a unit jump across the interface. The curvature can be calculated using $\kappa = \nabla \cdot \mathbf{n}$.
According to the continuous surface model (CSF) method~\cite{lafaurie1994_JCompPhys} and its remedies~\cite{morris2000_IntJNumerMethFluids,hu2006_adams_JCompPhys,wang2023_et_al_Frontiers}, the surface force can be written as a tensor form $\mathbf{F}^s = \nabla \cdot \Pi$, where the surface stress is computed as
\begin{eqnarray}
    \Pi = \alpha(\frac{1}{3}\mathbf{E}-\mathbf{n}\mathbf{n}),
\end{eqnarray}
where $\mathbf{E}$ is the identity matrix.


\subsection{Smoothed particle dynamics}
For convenience, we define some simple notations as reference
\begin{eqnarray}
        \mathbf{r}_{i j} &=&\mathbf{r}_{i}-\mathbf{r}_{j}, \\
        \mathbf{v}_{i j} &=&\mathbf{v}_{i}-\mathbf{v}_{j}, \\
        \mathbf{e}_{i j} &=&\mathbf{r}_{i j} / r_{i j}, \quad r_{i j}=\left|\mathbf{r}_{i j}\right|,
\end{eqnarray}
where $\mathbf{r}_i$, $\mathbf{v}_i$ are position and velocity of SPD particle i; 
$\mathbf{r}_{ij}$, $\mathbf{v}_{ij}$ are relative position and velocity of particles $i$ and $j$; $r_{ij}$ is the distance of the two and $\mathbf{e}_{ij}$ is the unit vector pointing from $j$ to $i$. 
Each particle's position is updated according to
\begin{eqnarray}
        \frac{\mathrm{d}\mathbf{r}_i}{dt} = \mathbf{v}_i. \label{position_eq}
\end{eqnarray}
The density field is computed as~\citep{espanol2003smoothed}
\begin{equation}
    \begin{aligned}
        \sigma_{i} &=\frac{\rho_{i}}{m_{i}}=\sum_{j} W\left(r_{i j}\right)=\sum_{j} W_{i j}\label{sigma_eq},
    \end{aligned}
\end{equation}
where $\sigma$ is number density defined as the ratio of $\rho$ and particle mass $m$ (constant).
Note that the density summation in Eq.~(\ref{sigma_eq}) together with the position update in Eq.~(\ref{position_eq}) already account for the continuity equation in Eq.~(\ref{continuity_eq}), which does not need to be discretized separately~\cite{espanol2003smoothed}.
The weight function $W(r)$, also known as kernel, has at least two properties:
    
\begin{subequations}
\label{weight_fun}
\begin{eqnarray}
\lim _{h \rightarrow 0} W\left(\mathbf{r}-\mathbf{r}^{\prime},h\right)=\delta\left(\mathbf{r}-\mathbf{r}^{\prime}\right)
\end{eqnarray}
\begin{equation}
\int W\left(\mathbf{r}-\mathbf{r}^{\prime}, h\right) d \mathbf{r}^{\prime}=1,
\end{equation}
\end{subequations}
where $h$ is quoted as smoothing length.
This indicates that any kernel adopted should converge to the Dirac delta function $\delta$ as $h\to0$ and its integral must be normalized.  To balance the computational efficiency and accuracy, a finite support domain described by a cutoff radius $r_c$ is usually adopted. 
When two particles' distance is larger than $r_c$, $W(r_{ij} \geq r_c,h)=0$ and there is no direct contribution to each other's dynamics. 
In this work we adopt the quintic spline function with $r_c = 3h$, which has been proven to be accurate~\citep{Morris1997_et_al_JCompPhys}:
\begin{widetext}
\begin{equation}
    \begin{aligned}
W(s, h)=C_d\frac{1}{h^d}\left\{
    \begin{array}{ll}
(3-s)^{5}-6(2-s)^{5}+15(1-s)^{5}, & 0 \leq s<1 ; \\
(3-s)^{5}-6(2-s)^{5}, & 1 \leq s<2 ; \\
(3-s)^{5}, & 2 \leq s<3 ; \\
0, & s \geq 3 .
    \end{array}\right.
\end{aligned} \label{quintic_function}
\end{equation}
\end{widetext}
Here $s = {r_{ij}}{/h}$ and $d$ is the number of dimension. The normalization coefficients are $C_2 = 7/478\pi$, and $C_3 = 3/359\pi$ in two and three dimensions, respectively. We use $h=1.2\Delta x$, where $\Delta x$ is the distance between initial neighboring particles. 
The squirmer particles are initially on a spherical coordinate system and other particles are on the cubic lattice.

The momentum equation of every SPD particle can be expressed succinctly as follows
\begin{eqnarray}
    m_i\frac{{\mathrm{d}}\mathbf{v}_i}{{\mathrm{d}}t} = \sum_j(\mathbf{F}_{ij}^C + \mathbf{F}_{ij}^D + \mathbf{F}_{ij}^A + \mathbf{F}_{ij}^R + \mathbf{F}_{ij}^S).
\label{SPD_eq}
\end{eqnarray}
Here $\mathbf{F}_{ij}^C$ and $\mathbf{F}_{ij}^D$ are conservative and dissipative forces between a pair of adjacent particles, the sum of which corresponds to a discretization of the forces due to pressure and viscous stress in the Navier-Stokes equations in Eq.~(\ref{ns_eq}). $\mathbf{F}_{ij}^A$ is the additional term to minimise numerical errors due to irregular distributions of particles~\cite{adami2013_JCompPhys} and appears only in macroscopic flow problems. $\mathbf{F}_{ij}^R$ is the random force used for mesoscopic flow problems~\cite{espanol2003smoothed}. $\mathbf{F}_{ij}^S$ represents the surface force acting only on the surface in multiphase flow.  There are a variety of formulations for the pairwise forces with different combinations~\cite{adami2013_JCompPhys, espanol2003smoothed, bian2012_et_al_PhysFluid, hu2006_adams_JCompPhys}. The proposed squirmer model is not restricted to any particular force formulation. 
In this work, we propose a unified discrete approach that can address the complexity of fluid-solid coupling at multiple scales and interface tracking for multiphase flows, while ensuring angular momentum conservation.

The conservative and dissipative terms are discretised as follows
\begin{eqnarray}
\mathbf{F}_{ij}^{C} &=&-(\frac{1}{\sigma_{i}^{2}}+\frac{1}{\sigma_{j}^{2}})\frac{\rho_{j}p_{i}+\rho_{i}p_{j}}{\rho_{i}+\rho_{j}}\frac{\partial W}{\partial r_{ij}}\mathbf{e}_{ij}, \label{sph_pressure}\\
\mathbf{F}_{ij}^{D} &=& (d+2) \left(\frac1{\sigma_i^2}+\frac1{\sigma_j^2} \right)\frac{2\eta_i\eta_j}{\eta_i+\eta_j}\frac{\partial W}{r_{ij} \partial r_{ij}} \mathbf{e}_{ij} \cdot \mathbf{v}_{ij} \mathbf{e}_{ij}, \label{sph_dissipation}
\end{eqnarray}
where the particle-averaged pressure and viscosity are employed, which are suitable for handling multiphase problems. 

The additional term is discretized as 
\begin{eqnarray}
\mathbf{F}_{ij}^{A} &=& \frac{1}{2}(\frac{1}{\sigma_i^2}+\frac{1}{\sigma_j^2})(\mathbf{A}_i+\mathbf{A}_j)\cdot\frac{\partial{W}}{\partial{r_{ij}}}\mathbf{e}_{ij},\label{addition_eq}
\end{eqnarray}
where $\mathbf{A} = \rho \mathbf{v}(\mathbf{\tilde{v}}-\mathbf{v})$ s a tensor from the dyadic product of the two vectors. 
Furthermore, $\mathbf{v}$ is the velocity for momentum and force calculations, while $\mathbf{\tilde{v}}$ is the modified transport velocity for updating the position of each fluid particle. The discrete form of $\mathbf{\tilde{v}}$ is calculated as
\begin{eqnarray}
\tilde{\mathbf{v}}_i = \mathbf{v}_i(t) &+& \delta t(\frac{d\mathbf{v}_i}{dt} -\frac{\chi}{m_i}\sum_j(\frac{1}{\sigma_i^2}+\frac{1}{\sigma_j^2})\frac{\partial{W}}{\partial{r_{ij}}}\mathbf{e}_{ij}).
\end{eqnarray}
where $\delta t$ is the time step, and the positive constant $\chi$ only appears here but not in the momentum equation.

To have a local thermodynamic equilibrium at the mesoscopic scale, the pair of random stress and dissipative stress are inherently related and must follow the fluctuation-dissipation theorem. In a discrete setting, for a given expression of the dissipative force $\mathbf{F}^D_{ij}$, 
we may resort to the GENERIC framework to obtain the corresponding random force $\mathbf{F}^R_{ij}$:
\begin{equation}
    \mathbf{F}_{ij}^{R}=\left(-\frac{8(d+2)\eta_i \eta_j k_BT}{\eta_i+\eta_j}(\frac{1}{\sigma_i^2}+\frac{1}{\sigma_j^2})\frac{\partial{W}}{r_{ij}\partial{r_{ij}}}\right)^{1/2}d\overline{\mathscr{W}}_{ij}\cdot \mathbf{e}_{ij},\label{frandom_eqn}
\end{equation}
where $k_B$ is the Boltzmann constant and $T$ is the temperature. $d\mathscr{W}$ is a matrix of independent increments of the Wiener process, and $d\overline{\mathscr{W}}$ is the symmetric part of it
\begin{equation}
    d\overline{\mathscr{W}}_{ij} = (d\mathscr{W}_{ij}+d\mathscr{W}_{ij}^T)/2.
\end{equation}
Furthermore, the following symmetry between particles $i$ and $j$ is preserved
\begin{equation}
    d\mathscr{W}_{ij} = d\mathscr{W}_{ji}.
\end{equation}
The independent increments of the Wiener process satisfy the following mnemotechnical $\mathrm{It\hat{o}}$ rules
\begin{equation}
    d\mathscr{W}_{ij}^{\alpha \beta}d\mathscr{W}_{kl}^{\kappa \lambda} = [\delta_{ik}\delta_{jl}+\delta_{il}\delta_{jk}]\delta^{\alpha \kappa}\delta^{\beta \lambda}dt.
\end{equation}
In case of single-phase flow, the dissipative force and random force are consistent with the version of the angular momentum formula of the Ellero and Espa{\~n}ol~\cite{ellero2018everything}.

In addition, the surface force is generated as follows~\cite{hu2006_adams_JCompPhys}

\begin{eqnarray}
    \mathbf{F}_{ij}^{S}=(\frac{\Pi_i}{\sigma_i^2} + \frac{\Pi_j}{\sigma_j^2})\frac{\partial W}{\partial r_{ij}}\mathbf{e}_{ij}
\end{eqnarray}
where $\Pi$ is the total surface stress of the particle of phase $k$ from interacting with neighboring particles of other phases $l$:
\begin{eqnarray}
    \Pi = \sum_l \Pi_{kl}^{(1)}, l\neq k.
\end{eqnarray}
and the $k-l$ phase interface stress is
\begin{eqnarray}
    \Pi_{kl}^{(1)} = \frac{\alpha^{kl}}{|\nabla C^{kl}|}\left( \frac{\mathbf{E} |\nabla C^{kl}|^2}{3} - \nabla C^{kl}\nabla C^{kl}\right)
\end{eqnarray}
Here, $\alpha^{kl}$ is surface tension coefficient between phase $k$ and $l$. The gradient of a color index $C$ can be obtained as
\begin{equation}
    \nabla C_i^{kl} = \sigma_i\sum_j(\frac{C_i^l}{\sigma_i^2} + \frac{C_j^l}{\sigma_j^2})\frac{\partial W}{\partial r_{ij}}\mathbf{e}_{ij}.
\end{equation}


\subsection{Modeling squirmers using SPD}
We employ SPD boundary particles to represent any solid body, including the squirmer. Once the boundary particles have been initialized, the boundaries are accurately delineated throughout the simulation.
SPD boundary particles possess identical mass and resolution $ \Delta x $ as the fluid particles, ensuring that a fluid particle in proximity to the boundary has a complete support domain. 
To satisfy the correct pressure gradient near the boundary, the pressure of each boundary particle is interpolated from the surrounding fluid particles~\cite{adami2012_et_al_JCompPhys}.

To obtain the flow field generated by the squirmer, we modify the dissipative forces between the fluid and solid particles at the boundary, altering the velocity of the fluid near the boundary to induce effective slip. We assume that the tangential velocity within the solid adjacent to the interface is distributed linearly. Since the slip velocity at each point on the interface is known, the task at hand is to calculate this artificial velocity of the solid particle.

\begin{figure}[ht]
    \centering  
    \begin{subfigure}{0.48\textwidth}  
        \centering  
        \includegraphics[width=\textwidth]{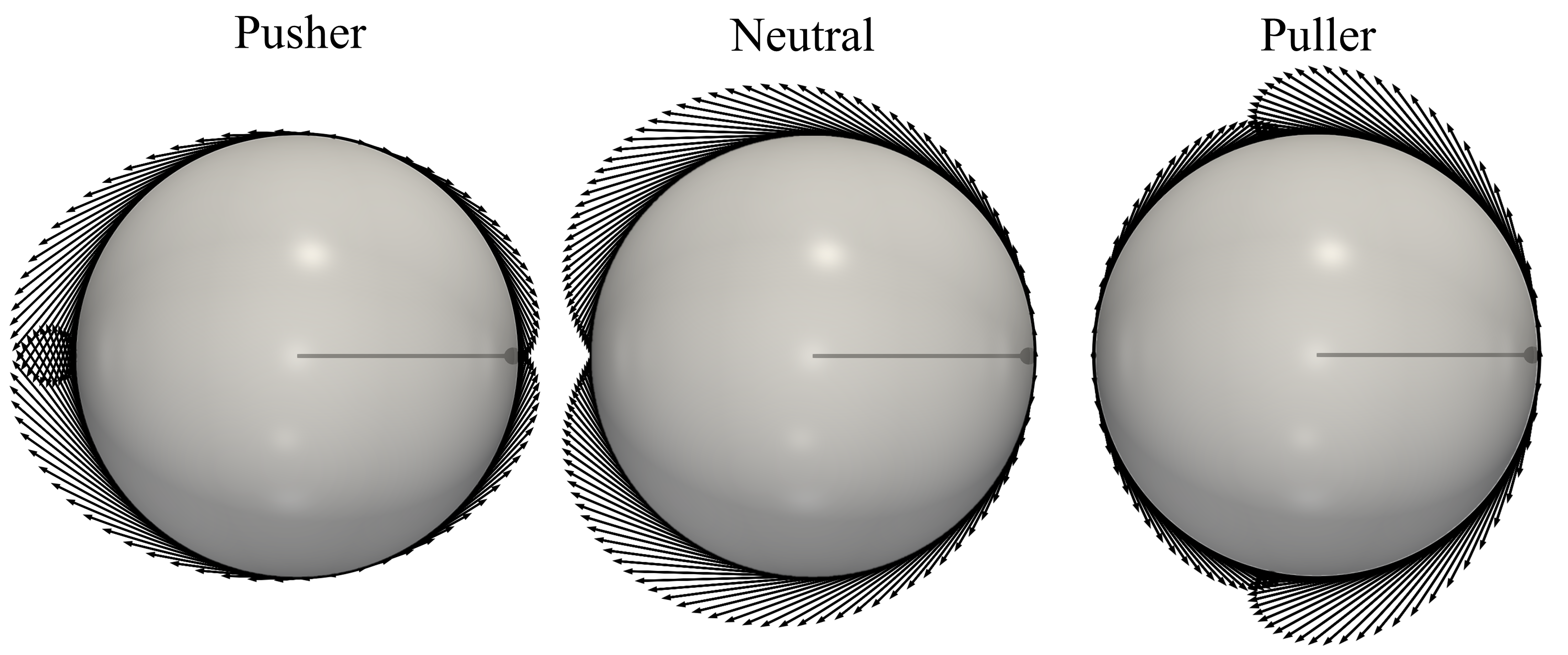}  
        \caption{Squirmer's slip velocity distribution at surface.}  
        \label{fig:Schematic_a}  
    \end{subfigure}
    \begin{subfigure}{0.5\textwidth}  
        \centering  
        \includegraphics[width=\textwidth]{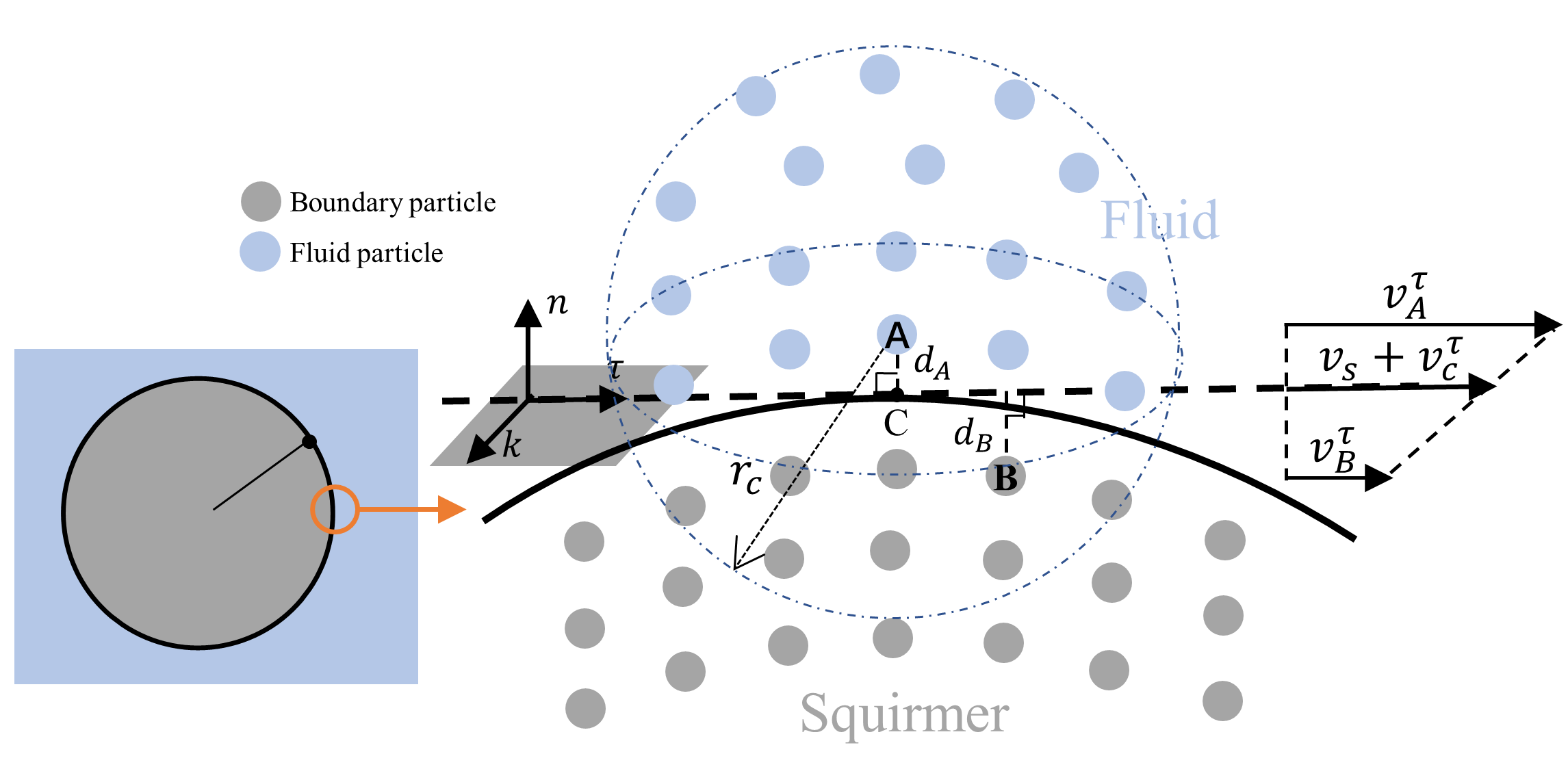}  
        \caption{Implementation of artificial velocities for boundary particles in squirmer.}  
        \label{fig:Schematic_b}  
    \end{subfigure}  
    \caption{Schematic of modeling a squirmer. To achieve the slip velocity of the squirmer in (a), artificial velocities of the boundary particles are assigned in (b). Given a fluid particle $A$, an interface plane tangential to the sphere is defined to be perpendicular to a line $AC$ of length $d_A$, where $C$ is the point of intersection on the surface. A Cartesian coordinate is chosen such that the direction $n$ is perpendicular to the interface plane, the direction $\tau$ is on the interface plane and coplanar with the plane containing $n$ and the direction of the squirmer head $\mathbf{e}$. $B$ is an example of neighbouring boundary particles with $d_B$ away from the interface plane. During the pairwise dissipative force calculation between $A$ and $B$, an artificial velocity $v_B$ is assigned to $B$ so that the linear interpolation between $v_A$ and $v_B$ has a tangential velocity $v_s+v_c^{\tau}$ at $C$ and the other directions keep the original velocity component. Given the same fluid particle $A$, all other nearby boundary particles follow the same procedure of $B$.}  
    \label{fig:Schematic}  
\end{figure} 

We define the normal distance from fluid particle $A$ to the sphere's boundary as $ d_A $, as depicted in Fig.~\ref{fig:Schematic}. This normal defines a tangent plane that is in contact with the sphere, which in two dimensions is a tangent line. The intersection of the normal with the tangent plane is denoted by point $C$. From this tangent plane, we can calculate the normal distance for each solid particle $B$ as $ d_B $. We can define a local coordinate system where $ n $ is the normal direction, and $ \tau $ and $ k $ lie within the tangent plane. In three dimensions, $ \tau $ and $ k $, which are perpendicular to each other, can be arbitrary within the tangent plane. However, for convenience, here $ \tau $ is coplanar with the plane containing $ n $ and the direction of the squirmer's head $\mathbf{e}$. The artificial velocity is assigned to the solid particles $B$ on the sphere, with the components in the normal and tangential directions specified as follows:



\begin{eqnarray}
v_B^{\tau} &=& -\frac{d_B}{d_A+\alpha h}(v_A^{\tau} - v_s - v_C^{\tau}) + v_s +v^{\tau}_C \nonumber \\
v_B^{n} &=& -\frac{d_B}{d_A+\alpha h}(v_A^n-v_C^n) + v_C^n \nonumber\\
v_B^{k} &=& -\frac{d_B}{d_A+\alpha h}(v_A^{k} - v_C^{k}) +v^{k}_C 
\label{artifical_velocity}
\end{eqnarray}
where $\mathbf{v}_B = (v_A^{\tau}, v_A^{n}, v_A^{k})$ is the fluid velocity of particle $A$. The surface slip velocity in tangential direction $v_s$ is defined in Eq.~(\ref{vs_theta}) or Eq.~(\ref{surface_velocity}). The $\alpha h$ term is included to keep the denominator nonzero and a
typical choice is $\alpha = 0.05$. $\mathbf{v}_C=(v_C^{\tau},v_C^{n}, v_C^k)$ represents the velocity of the intersection point on the surface
\begin{equation}
\mathbf{v}_C = \mathbf{\Omega}_0 \times \mathbf{r}_s + \mathbf{V}_0.
\end{equation}

Since the slip velicity $\mathbf{v}_s(\mathbf{r}_s, \mathbf{e})$ in Eq.~(\ref{surface_velocity}) only appears in the tangential direction, Eq.~(\ref{artifical_velocity}) can be written as
\begin{equation}
    \mathbf{v}_B = -\frac{d_B}{d_A+\alpha h}\left(\mathbf{v}_A-\mathbf{v}_s(\mathbf{r}_s, \mathbf{e})-\mathbf{v}_C\right) + \mathbf{v}_s(\mathbf{r}_s, \mathbf{e})+\mathbf{v}_C
\end{equation}
Then the relative velocity $\mathbf{v}_{AB}$ between $A$ and $B$ can be directly involved in computing the parawise dissipative force
\begin{eqnarray}
    \mathbf{v}_{AB} = \frac{d_A+\alpha h + d_B}{d_A + \alpha h}(\mathbf{v}_A-\mathbf{v}_s(\mathbf{r}_s, \mathbf{e})-\mathbf{v}_C)
\end{eqnarray}


During the computation of the pairwise dissipative forces between particles $A$ and $B$, an artificial velocity $ \mathbf{v}_B $ is assigned to particle $B$. This ensures that the linear interpolation between $ \mathbf{v}_A $ and $ \mathbf{v}_B $ satisfies the absolute velocity condition at the intersection point $C$, as described by Eq.~(\ref{absolute_velocity}). For a given fluid particle $A$, any boundary particle within its support domain follows the same procedure for $B$. And for the same boundary particle $B$, the artificial velocities are different when it interacts with different fluid particles $A$.
The artificial velocity $\mathbf{v}_B$ of a boundary particle is employed only in the calculation of dissipative force $\mathbf{F}_{ij}^D$ in Eq.~(\ref{sph_dissipation}), but not intended for the kinematics of the squirmer.

\section{Results}\label{section_results}
To illustrate the efficacy of the proposed methodology for the squirmer model, we conduct a series of simulations, encompassing a range of scenarios, from relatively simple to highly complex. In the absence of thermal fluctuations, we test a single squirmer at steady state, and analyze the resulting flow field it generates. Subsequently, we examine the hydrodynamic interactions between a squirmer and a wall, as well as between two squirmers. Afterwards, we test the dynamics of a squirmer when thermal fluctuations of the fluid are present. Finally, to expand the versatility of the model, we consider a squirmer in a multiphase flow environment. Unless otherwise stated, the resolution of the SPD method is set such that the spatial discretization, denoted by $\Delta x$, is equivalent to  $R/10$. This means that there are $10$ discrete particles uniformly distributed along the radius of the sphere. The resolution study in the first subsection demonstrates that this resolution is sufficient, ensuring both accuracy and computational efficiency.

\subsection{A single squirmer}
First, we study the motion of a single squirmer in fluid using SPD simulation.
The radius of the squirmer is taken to be unity ($R = 1$), which serves as the fundamental length scale for our simulations. The computational domain is a cubic box with an edge length of $L = 30$, using periodic boundary conditions to mimic an effectively infinite fluid environment.
The dynamic viscosity and density of the fluid are set to $\eta = 1$ and $\rho = 1$. The Reynolds number of a squirmer in bulk is defined as 
\begin{eqnarray}
    Re =  \frac{\rho R U_0}{\eta} =\frac{2\rho R B_1}{3\eta}.
\end{eqnarray}

\begin{figure}[h]
\includegraphics[scale=0.7]{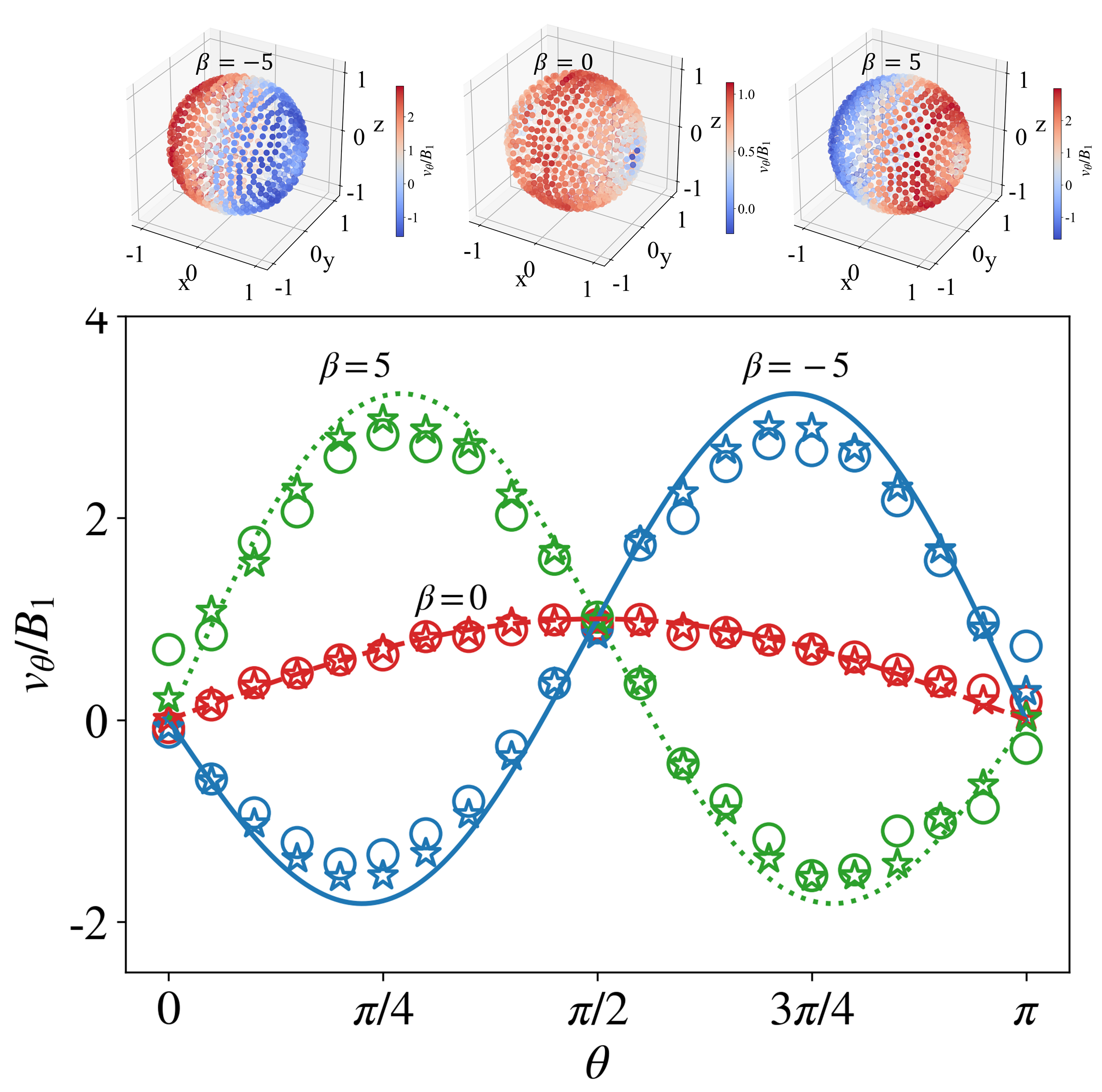}
\caption{\label{fig:surface_velocity} The distribution of the slip velocity  on the squirmer surface. The results of the SPD for $Re=0.01$ are expressed as scatter points. The circle represent a resolution of $\Delta x=R/10$, while the star denote $\Delta x=R/16$. The analytic solution for $Re=0$ is expressed as lines. Top are the distribution of fluid SPD particles and their slip velocity within one resolution from the squirmer. }
\end{figure}

\begin{figure}[h]
\includegraphics[scale=0.7]{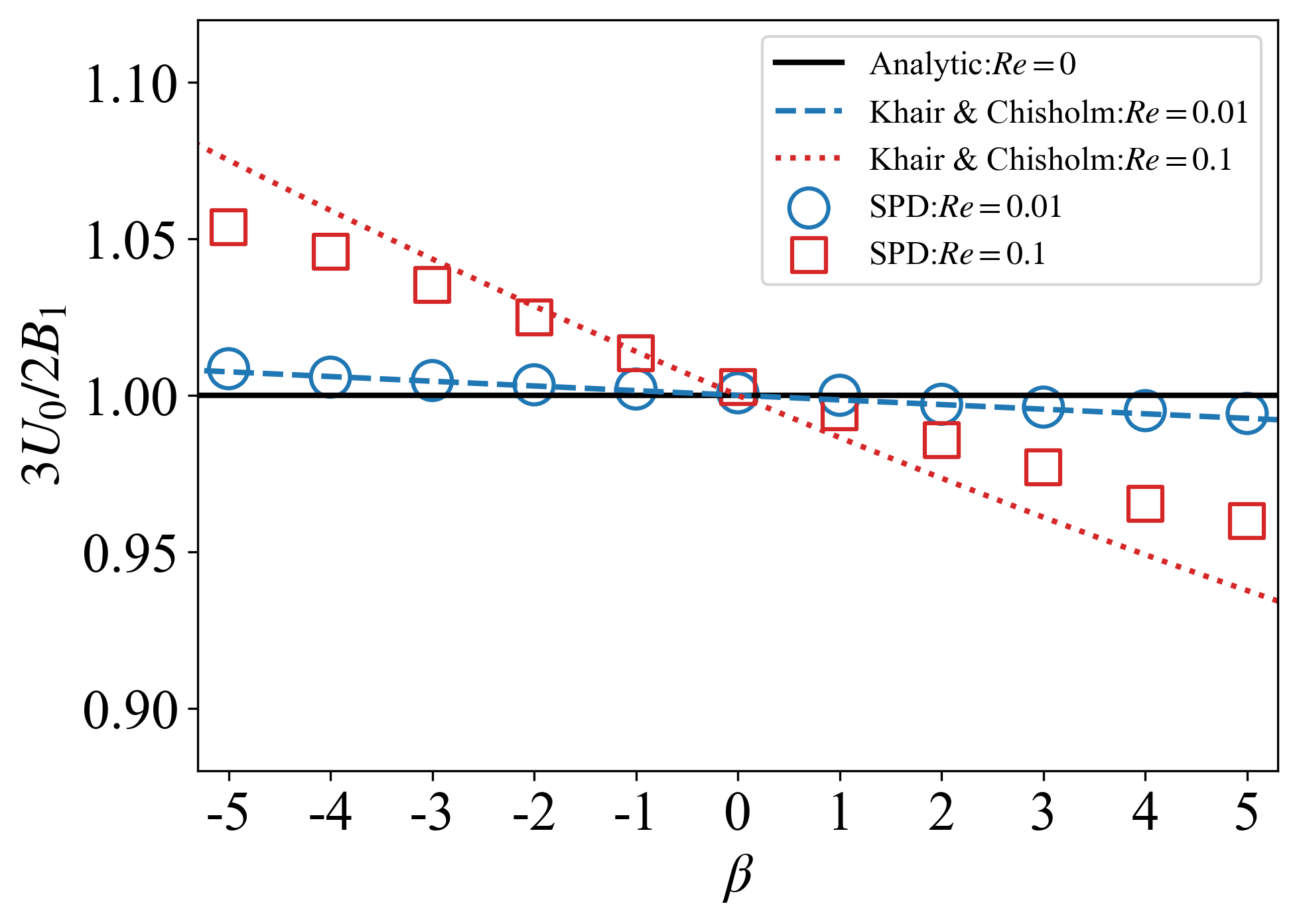}
\caption{\label{fig:variation_beta} The velocity magnitude of the squirmer with different $\beta$. The black dashed line represents the analytical velocity solution at zero Reynolds number. The dashed line results are from Eq.~(40) in Khair\&Chisholm~\cite{khair2014_Chisholm_PhysFluid}.}
\end{figure}

We first extract the surface slip velocity generated by a single squirmer at steady state. The squirmers are characterized by the first modes $B_1=0.015$, which is equivalent to a Reynolds number of $0.01$. Fig.~\ref{fig:surface_velocity} illustrates the slip velocity distributions for three distinct types of squirmers. The upper part of the figures indicate the slip velocity distribution of a fluid particle within one SPD resolution ($\Delta x$) from the squirmer. The lower part shows the variation of slip velocity with polar angle after interpolating the fluid particles to the surface of the sphere by the quintic spline kernel function in Eq.~(\ref{quintic_function}). The results of the simulation are in good agreement with the analytic solution for zero Reynolds number. 
We then check the steady state velocities $U_0$ of the squirmer for different values of squirmer parameter $\beta$,  as shown in Fig.~\ref{fig:variation_beta}. Two groups of squirmer's first modes are chosen, $B_1=0.015$ and $B_1=0.15$, corresponding to Reynolds numbers of $Re=0.01$ and $Re=0.1$, respectively. The black dashed line represents the analytical velocity solution at zero Reynolds number, where the steady-state velocity is independent of $\beta$, i.e. $3U_0/2B_1=1$. The dashed lines indicate the analytical results obtained using perturbation theory~\cite{khair2014_Chisholm_PhysFluid}. For pusher, both the simulation and the perturbation theory are slight larger than the analytic solution for zero Reynolds number, while for pullers the results are reversed. This implies that pushers accelerate and pullers decelerate in the presence of inertia. The maximum relative error between the our results and the perturbation theory is $0.1\%$ for $Re=0.01$ and $2\%$ for $Re=0.1$ at $\beta=5$.


\begin{figure}[h]
\includegraphics[scale=0.6]{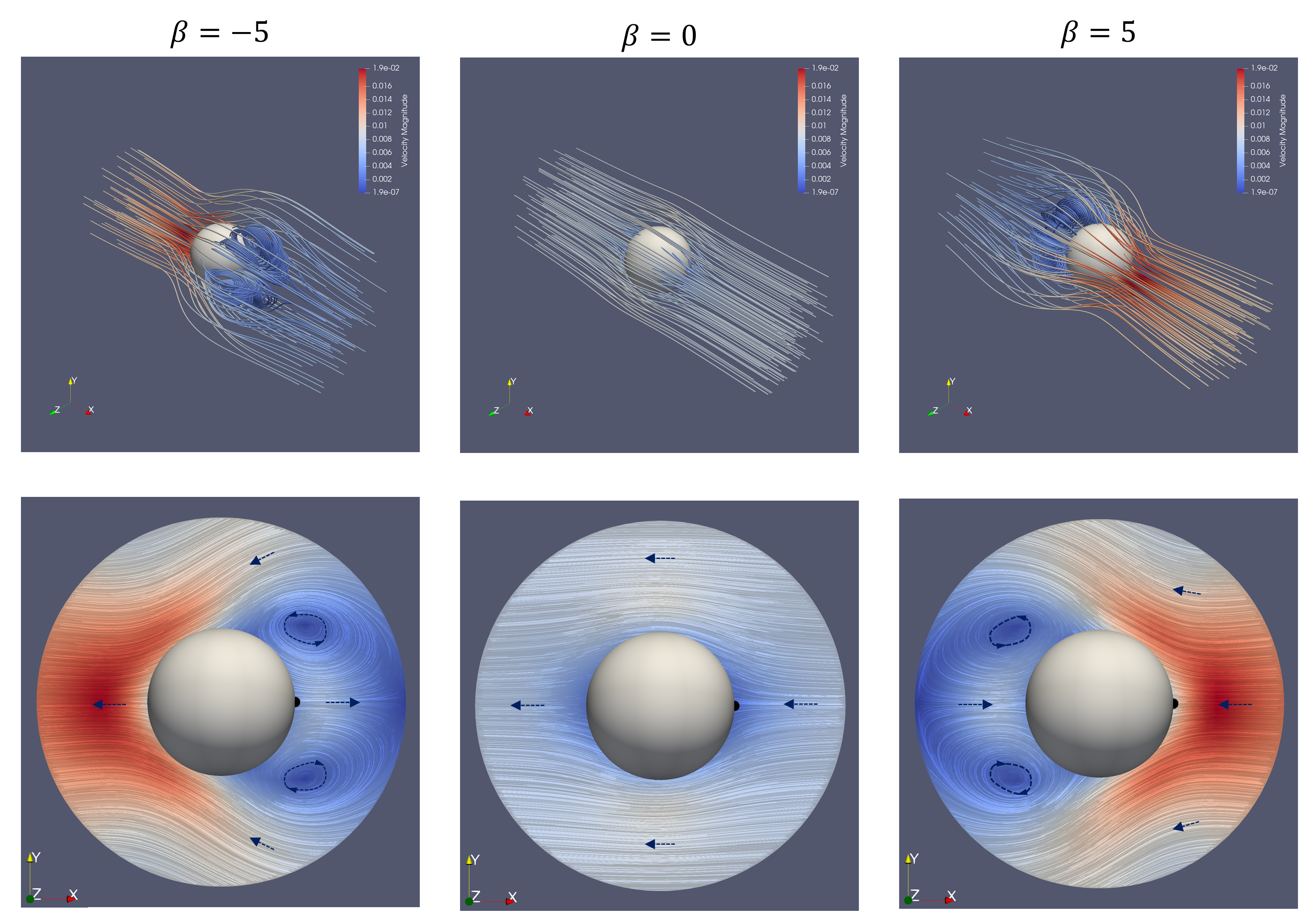}
\caption{\label{fig:streamline} The streamlines for a pusher ($\beta=-5$), a neutral squirmer ($\beta=0$) and a puller ($\beta=5$ ) in the frame moving with the swimmer. The swimmers swim forward along the $x$-axis. The upper part shows the streamlines in 3D and the bottom part shows a slice. The arrows indicate the direction of certain stream lines.}
\end{figure}

Fig.~\ref{fig:streamline} shows the velocity field generated by a pusher ($\beta=-5$), a neutral swimmer ($\beta=0$) and a puller ($\beta=5$) in the frame moving with the swimmers. All swimmers move in the positive direction of the $x$-axis. The first row shows the 3D streamlines and the second row shows the streamlines in a tangent plane perpendicular to the $z$-axis and over the centre of the sphere. The vortexes generated by the pusher is in front of the swimming direction, while the vortexes generated by the puller is behind the swimming direction. Neutral swimmers do not generate any vortexes. 

\begin{figure*}[h]
    \centering  
    \begin{subfigure}{0.32\textwidth}  
        \centering  
        \includegraphics[width=\textwidth]{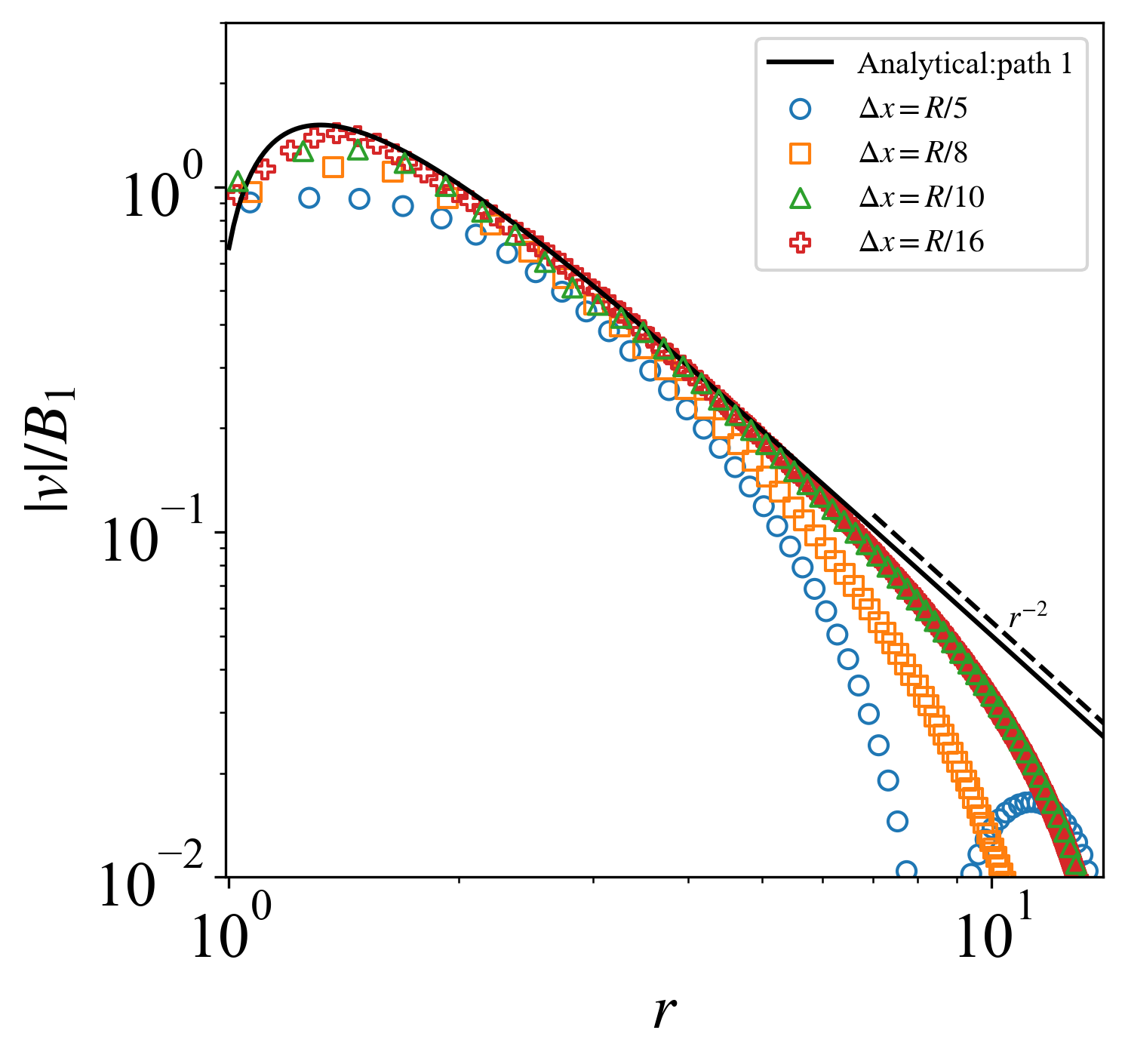}  
        \caption{Path 1}  
        \label{fig:sub1}  
    \end{subfigure}  
    \begin{subfigure}{0.32\textwidth}  
        \centering  
        \includegraphics[width=\textwidth]{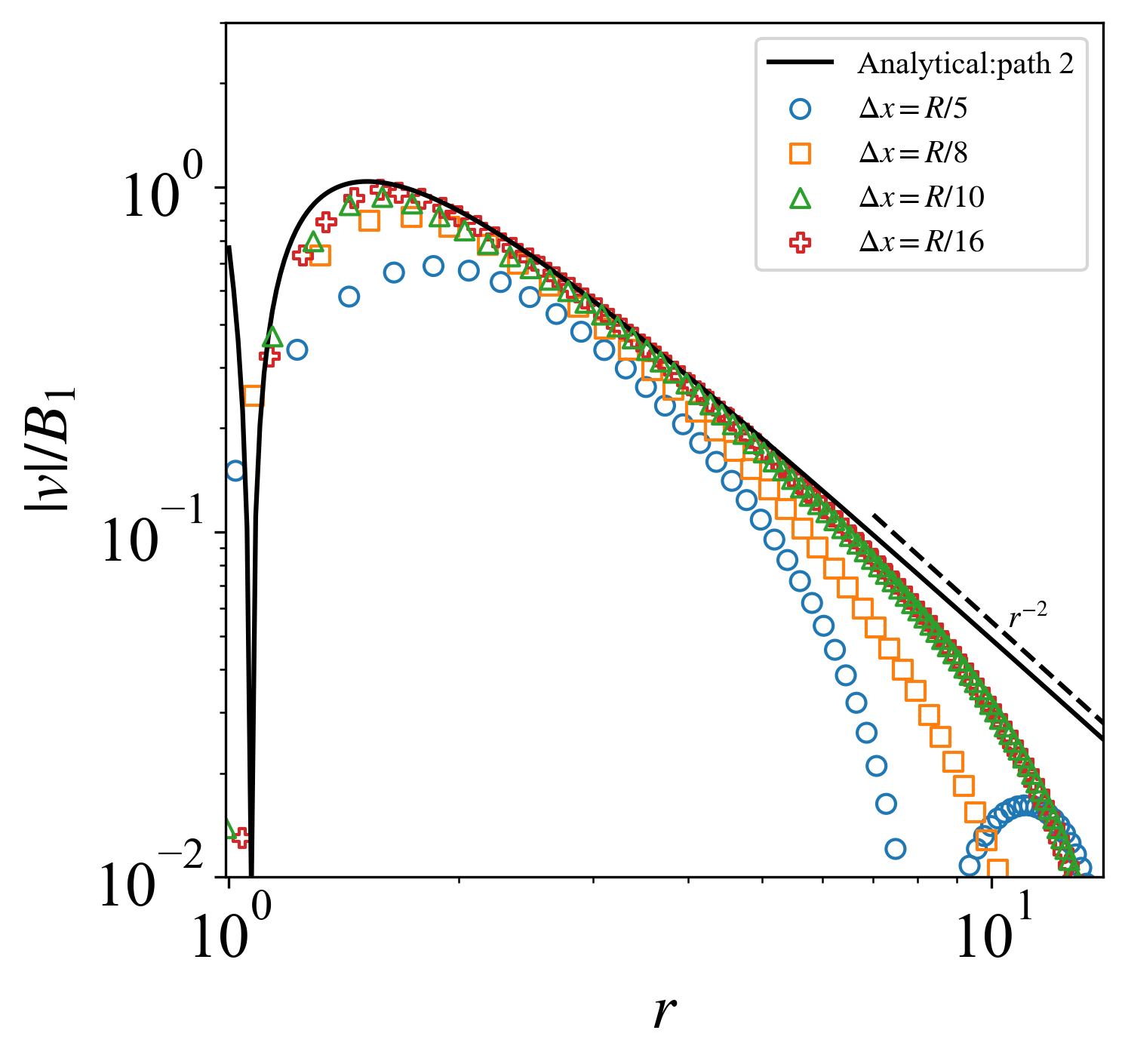}  
        \caption{Path 2}  
        \label{fig:sub2}  
    \end{subfigure}  
    \begin{subfigure}{0.32\textwidth}  
        \centering  
        \includegraphics[width=\textwidth]{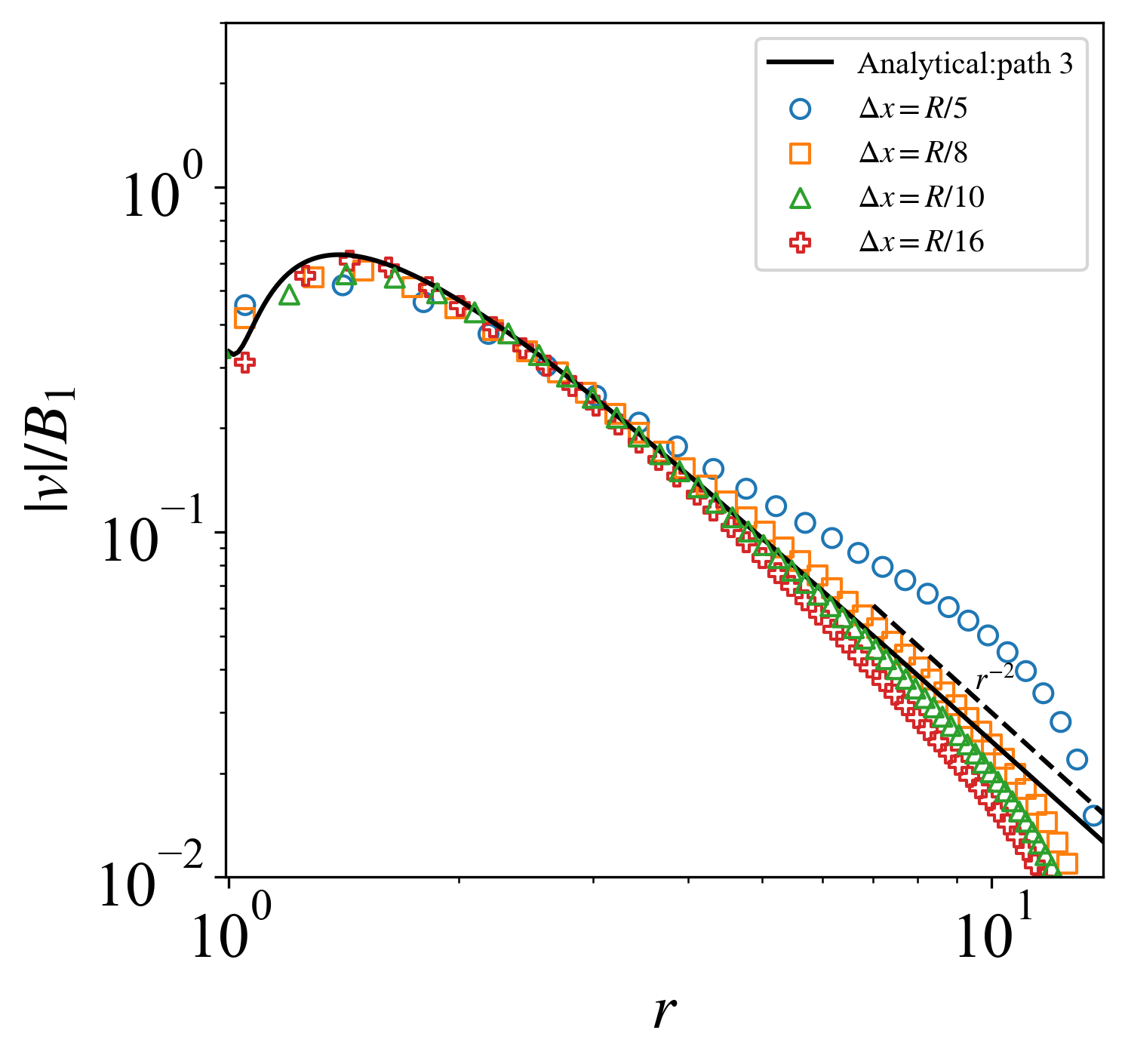}  
        \caption{Path 3}  
        \label{fig:sub3}  
    \end{subfigure}  
    \caption{Decay of the flow field. $|v|$ is the velocity magnitude of the fluid. (a) Path 1 is along the negative swimming direction, (b) path 2 is along the positive flow direction and (c) path 3 is perpendicular to the flow direction. Black solid lines represent the analytic solution of the velocity filed around a squirmer in bulk.}  
    \label{fig:decay_flow_field}  
\end{figure*}

To analyse the flow field quantitatively, we compare the velocities in three directions generated by the swimmer with the theoretical solution in Eq.~(\ref{velocity_field_squirmer}). These directions are path 1 along the negative swimming direction, path 2 along the positive flow direction and path 3 perpendicular to the flow direction. 
Fig.~\ref{fig:decay_flow_field} shows the velocity decay of the fluid field generated by a puller with the Reynolds number of $Re=0.01$ at different SPD resolutions. The black solid line is the theoretical solution in Stokes flow, as detailed in Appendix~\ref{Velocity_field_analytic}, while the scattered points illustrate the results of the SPD simulation. Resolution study shows that our results converge. And the high resolution results ($\Delta x = R/16$) agree with the theoretical solution. The squirmer in the far field is consistent with an $r^{-2}$ decay. Further away the decay is faster due to periodic effects. This effect can be reduced by increasing the simulation domain. See box size study in Appendix~\ref{appendix_box_size}. According to the results of the resolution study, although the high resolution of $\Delta x=R/16$ gives better results, considering the simulation accuracy as well as the computational efficiency, it is sufficient to distribute $10$ SPD particles on the radius in the 3D simulation.

\subsection{Hydrodynamic interactions}
The microorganisms may occur in multiples and are usually subject to geometric constraints. Hydrodynamic interactions between microorganisms and between microorganisms and boundaries play a crucial role. Next, we validate the pairwise interactions of squirmers as well as the dynamic behaviour of a squirmer near a wall. The radius of the squirmers are all $R = 1$. The dynamic viscosity and density of the fluid are set to $\eta = 1$ and $\rho = 1$. 




When multiple squirmers are in close proximity to each other or near a wall, the gaps between them are often small. The resolution of the simulation may fails to capture the dynamics within these gaps, potentially leading to unphysical SPD particle penetration. To address this issue, we introduce short-range repulsive forces~\cite{glowinski2001_JCompPhys, spagnolie2012_JFluidMech, li2014_PhysRevE}:

\begin{equation}
    \mathbf{f}_r=\frac{C_m}{\epsilon}\left(\frac{d-d_{\min}-dr}{dr}\right)^2 \mathbf{e}_r,
\end{equation}
where $C_m = \frac{MU_0^2}{R}$ is a scaling factor with the dimension of force, $\epsilon$ is a small positive number (typically $10^{-4}$), $d$ is the distance between two swimmers or the distance between the center of a squirmer and the wall, and $ d_{\min} = R $ or $ 2R $ is the corresponding minimum possible distance. 
$\mathbf{e}_r$ is the direction of the repulsive force along the line connecting the centres of the two squirmers or perpendicular to the wall. $dr$ is the range of the force and in this work is set to $ 0.5\Delta x$.

\begin{figure}  
    \centering  
    \begin{subfigure}{0.45\textwidth}  
        \centering  
        \includegraphics[width=\textwidth]{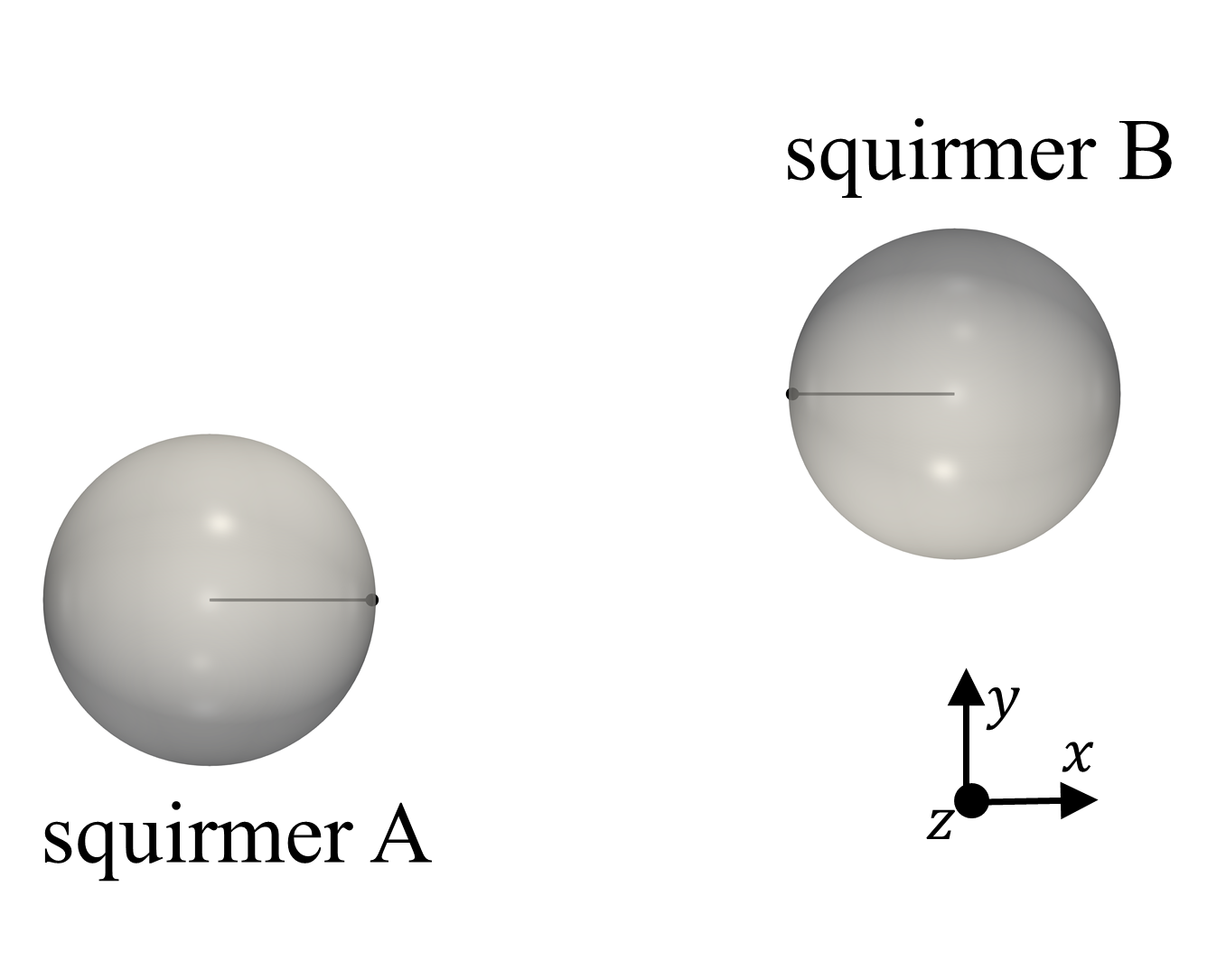}  
        \caption{Two squirmers interaction.}  
        \label{fig:interaction_a}  
    \end{subfigure}
    \begin{subfigure}{0.45\textwidth}  
        \centering  
        \includegraphics[width=\textwidth]{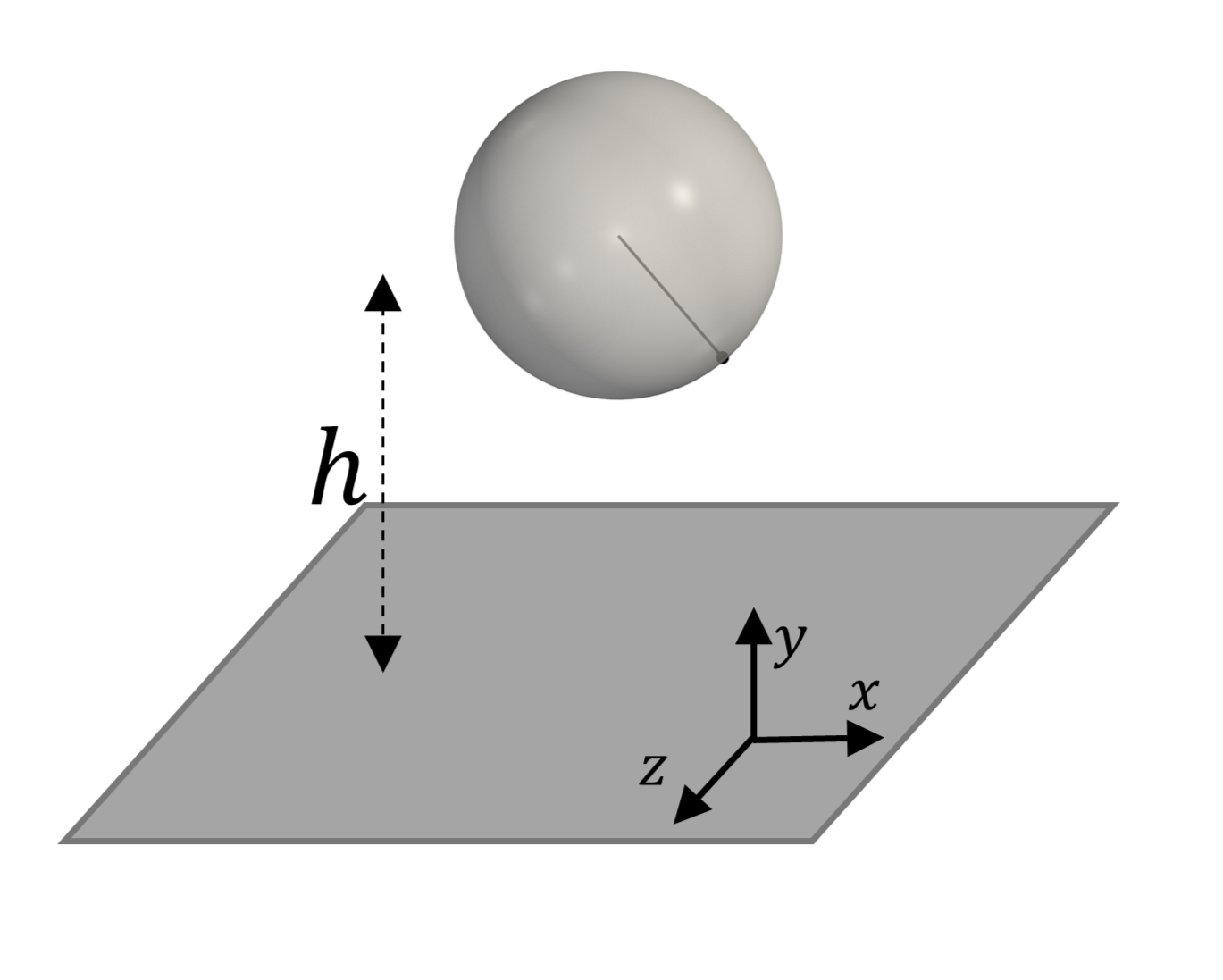}  
        \caption{A squirmer near a wall.}  
        \label{fig:interaction_b}  
    \end{subfigure}  
    \caption{Schematic of the hydrodynamic interaction of the squirmer(s) at starting point.}  
    \label{fig:interaction}  
\end{figure}  

\begin{figure}  
    \centering  
    \begin{subfigure}{0.45\textwidth}  
        \centering  
        \includegraphics[width=\textwidth]{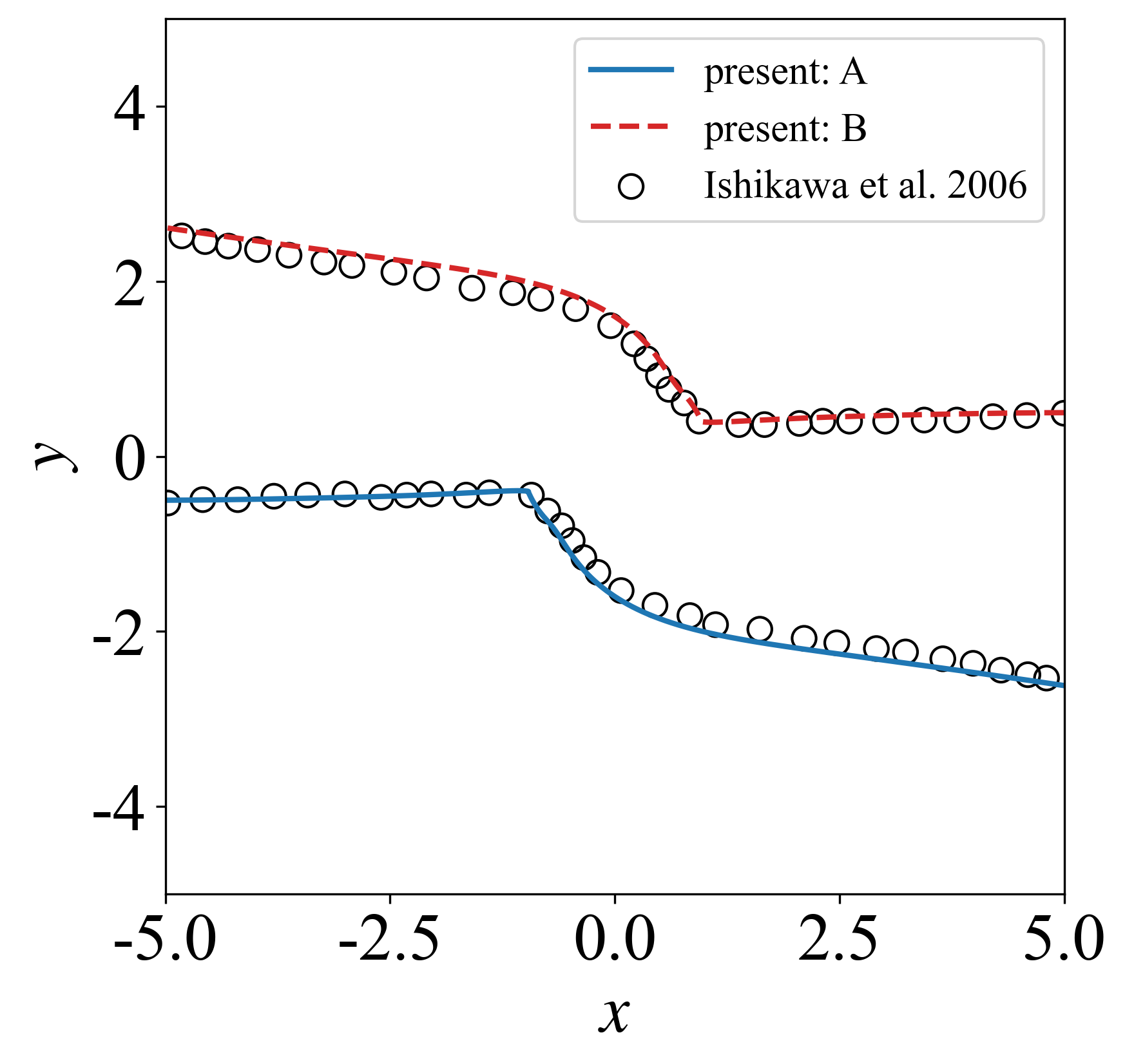}  
        \caption{$\beta=5$}  
    \end{subfigure} 
    \begin{subfigure}{0.45\textwidth}  
        \centering  
        \includegraphics[width=\textwidth]{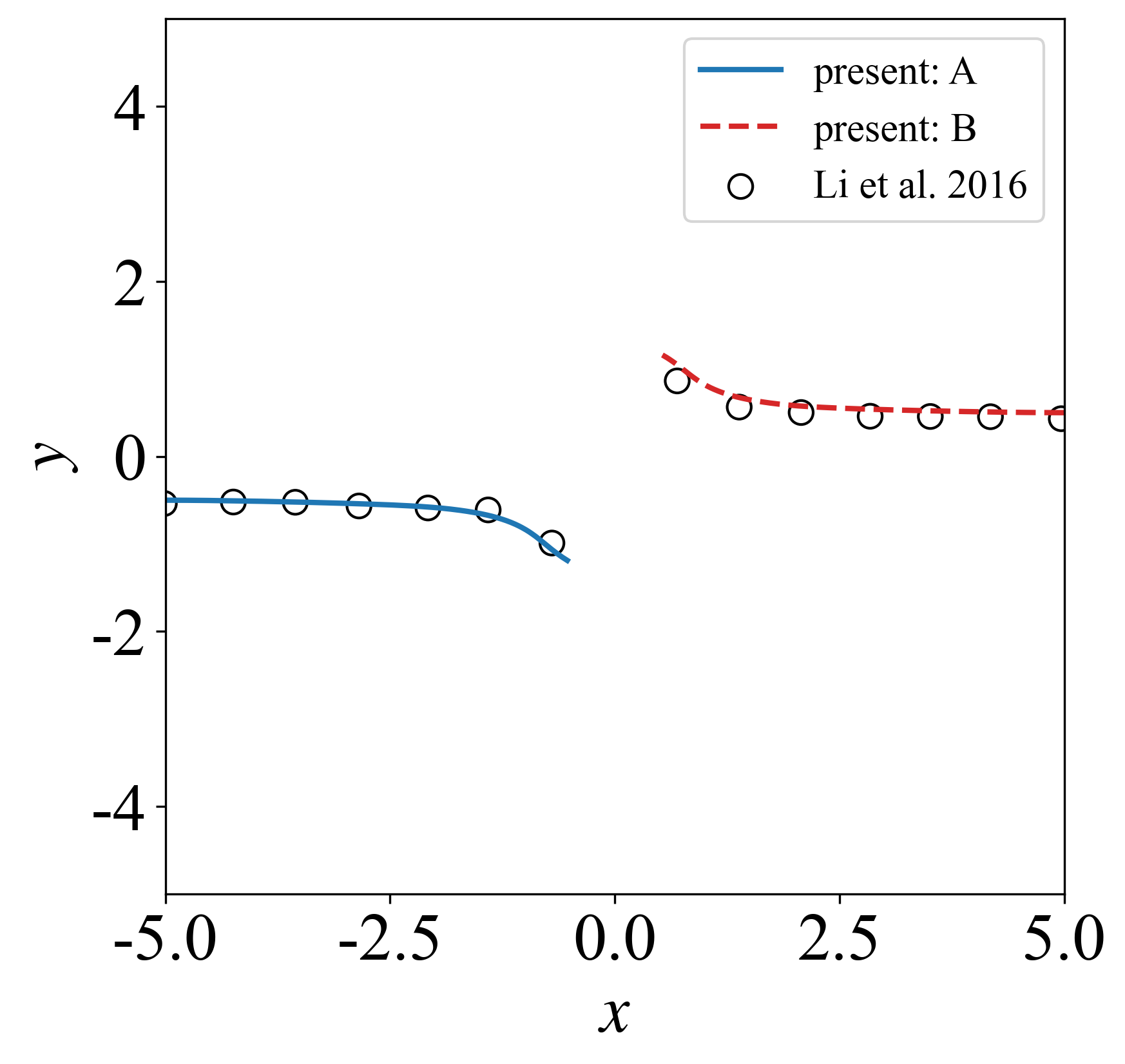}  
        \caption{$\beta=-5$}  
    \end{subfigure}  
    \caption{Comparison of the trajectories of two colliding microswimmers. }  
    \label{fig:2squirmers}  
\end{figure}

First, we simulate the mutual collision of two squirmers. A schematic of this simulation is shown in Fig.~\ref{fig:interaction_a}. The computational region is a cubic box with side length $L = 30$, surrounded by periodic boundary conditions. We consider the pairwise interactions of two pushers ($\beta = -5$) and two pullers ($\beta = 5$). 
The initial configuration of the two squirmers (represented by A and B) is that they are in the same plane ($z = 0$), with parallel head directions and facing each other, swimming with a velocity $U_0=0.1$. The corresponding Reynolds number is $Re=0.1$.  
The distances between their centers in the $x$- and $y$-directions are $dx=5$ and $dy=1$ respectively. The paths of the two squirmers are shown in Fig.~\ref{fig:2squirmers}. The lines represent the results of the present simulation, while the black scatter points represent the reference solution. For pushers with $\beta = 5$, our results are consistent with those of Ishikawa et al.~\cite{ishikawa2006_et_al_JFluidMech} in Stokes flow. The two squirmers remain in the plane and separate after approaching each other. For pullers with $\beta = -5$ our results agree with those of Li et al.~\cite{li2016_PhysRevE} where the two squirmers become entangled and leave the $z$ plane after approaching each other.

\begin{figure}  
    \centering  
    \begin{subfigure}{0.45\textwidth}  
        \centering  
        \includegraphics[width=\textwidth]{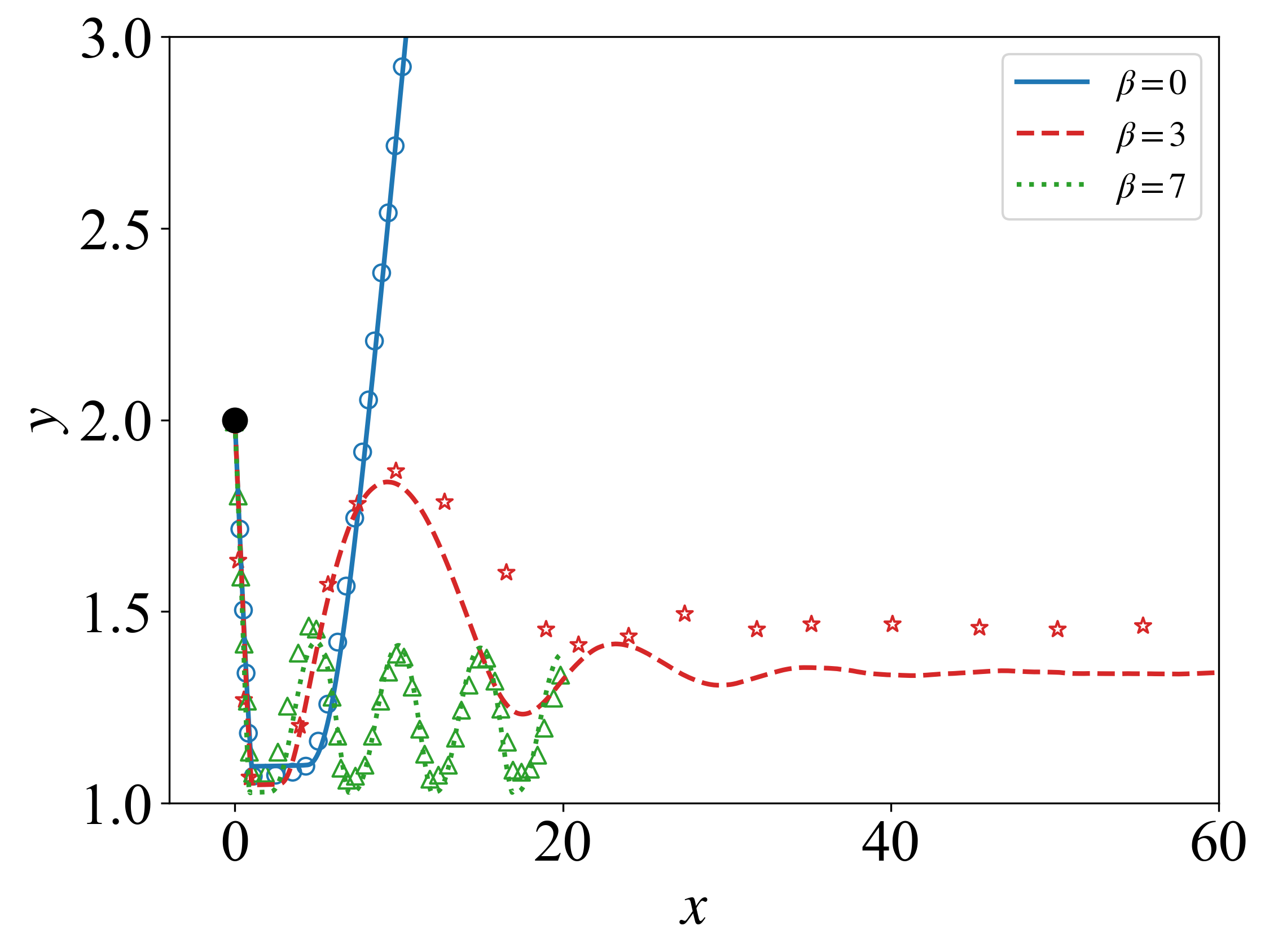}  
        \caption{Trajectory of the squirmer.}  
        \label{fig:trajectory}  
    \end{subfigure}
    \begin{subfigure}{0.45\textwidth}  
        \centering  
        \includegraphics[width=\textwidth]{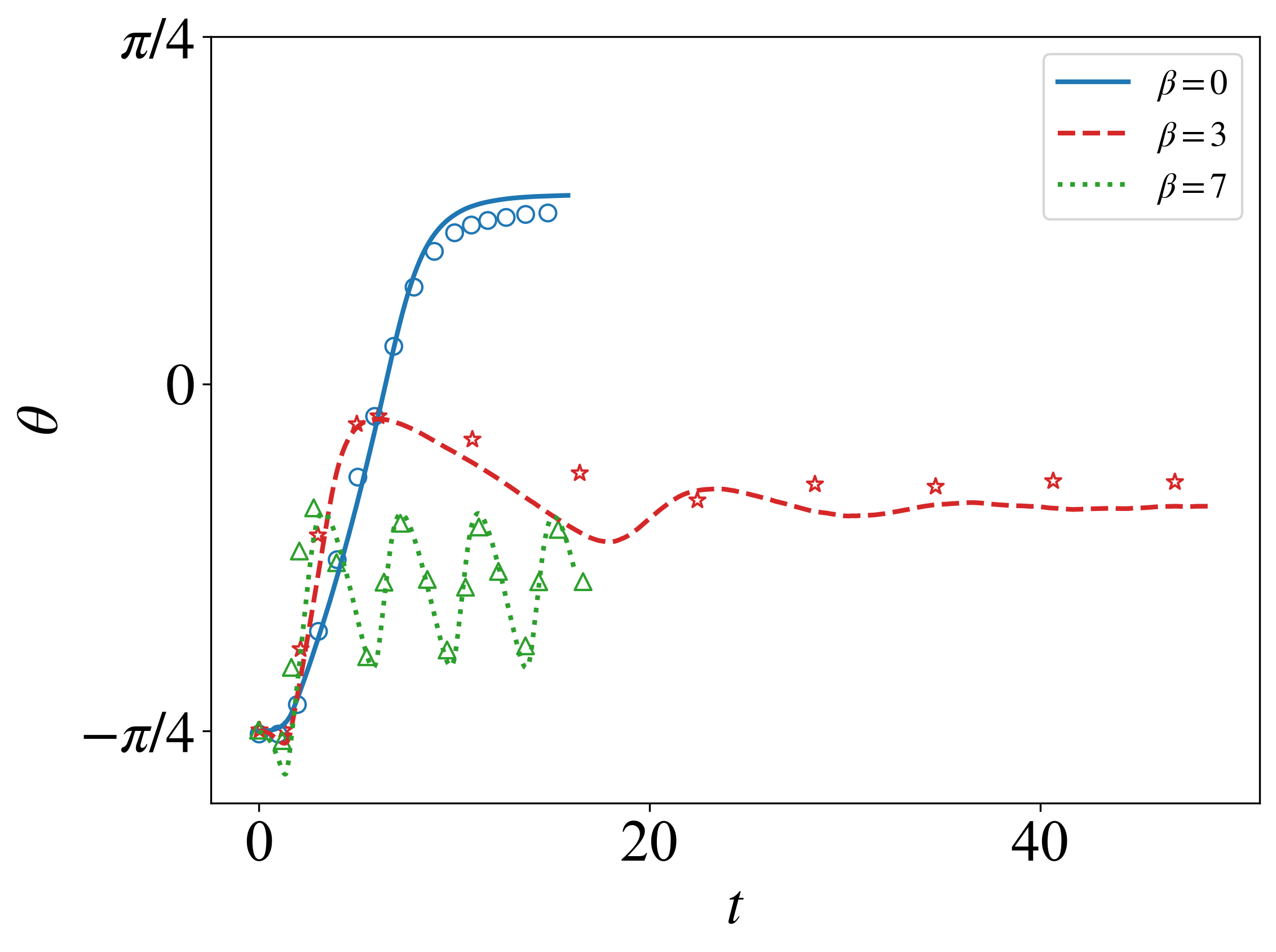} 
        \caption{Temporal evolution of orientation angle.}  
        \label{fig:orientation}  
    \end{subfigure}  
    \caption{Trajectory and temporal evolution of orientation angle $\theta$ for squirmers  at $Re = 1$. The squirmer initially is located at $h=2R$, $\theta=\pi/4$. The present results are shown as lines, the results of Li et al.~\cite{li2014_PhysRevE} as scatter.}  
    \label{fig:near_wall}  
\end{figure}

We then consider a squirmer swimming near a wall. Li et al.~\cite{li2014_PhysRevE} identified three types of motion for a single squirmer near a wall at a Reynolds number of $1$: (1) for $\beta \leq 1$, the squirmer swims away from the wall at a positive angle; (2) for $2\leq \beta \leq 5$, the squirmer oscillates near the wall and eventually swims along it at a constant distance and at a negative angle; (3) for $\beta \geq 7$, the squirmer bounces forward on the wall. We have verified these three types of motion separately, with $\beta$ chosen to be $0$, $3$ and $7$. 
Fig.~\ref{fig:interaction_b} shows a schematic of the squirmer near the wall at the initial time. The simulation region of the fluid is a cube with side length $L=30$, with periodic boundaries in the $x$ and $z$ directions and no-slip walls in the $y$ direction. The walls are composed of boundary particles. The squirmer is close to the lower wall, and the effect of the upper wall on the squirmer can be assumed to be negligible. The center of mass of the squirmer is at a height of $h = 2R$ from the lower wall, with its head swimming towards the wall at an angle of $-45^{\circ}$ degrees to the $x$ axis. Fig.~\ref{fig:trajectory} shows the trajectory of the squirmer's movement, with the black point as its starting point. Fig.~(\ref{fig:orientation}) shows the angle of the squirmer's head with respect to the $x$-direction over time. The lines are the results of our simulations and the scatter points are the reference solutions. We reproduce the three motion types of the squirmer with three different $\beta$. The trajectories and the temporal evolution of orientation angle are in good agreement with the results of Li et al. The trajectories of the squirmer with $\beta = 3$ diverge slightly from the reference solution in the second half of the simulation. This discrepancy may be attributed to the boundary conditions of our simulation region differing from those employed by Li et al.

\subsection{A mesoscale squirmer with thermal fluctuations}


In our investigation of the squirmer's dynamics at the mesoscopic scale, we establish a scenario where the thermal energy is quantified by the product of the Boltzmann constant and temperature, $k_BT = 1$. Considering the computational efficiency, the simulation is conducted within a cubic domain with side length $L = 15$ along each side, with periodic boundary conditions applied to encapsulate the system. The squirmer is characterized by a unit radius, $R = 1$, navigating through the fluid with density of $\rho = 1$ and dynamic viscosity of $\eta = 15$.

\begin{figure}  
    \centering  
    \begin{subfigure}{0.45\textwidth}  
        \centering  
        \includegraphics[width=\textwidth]{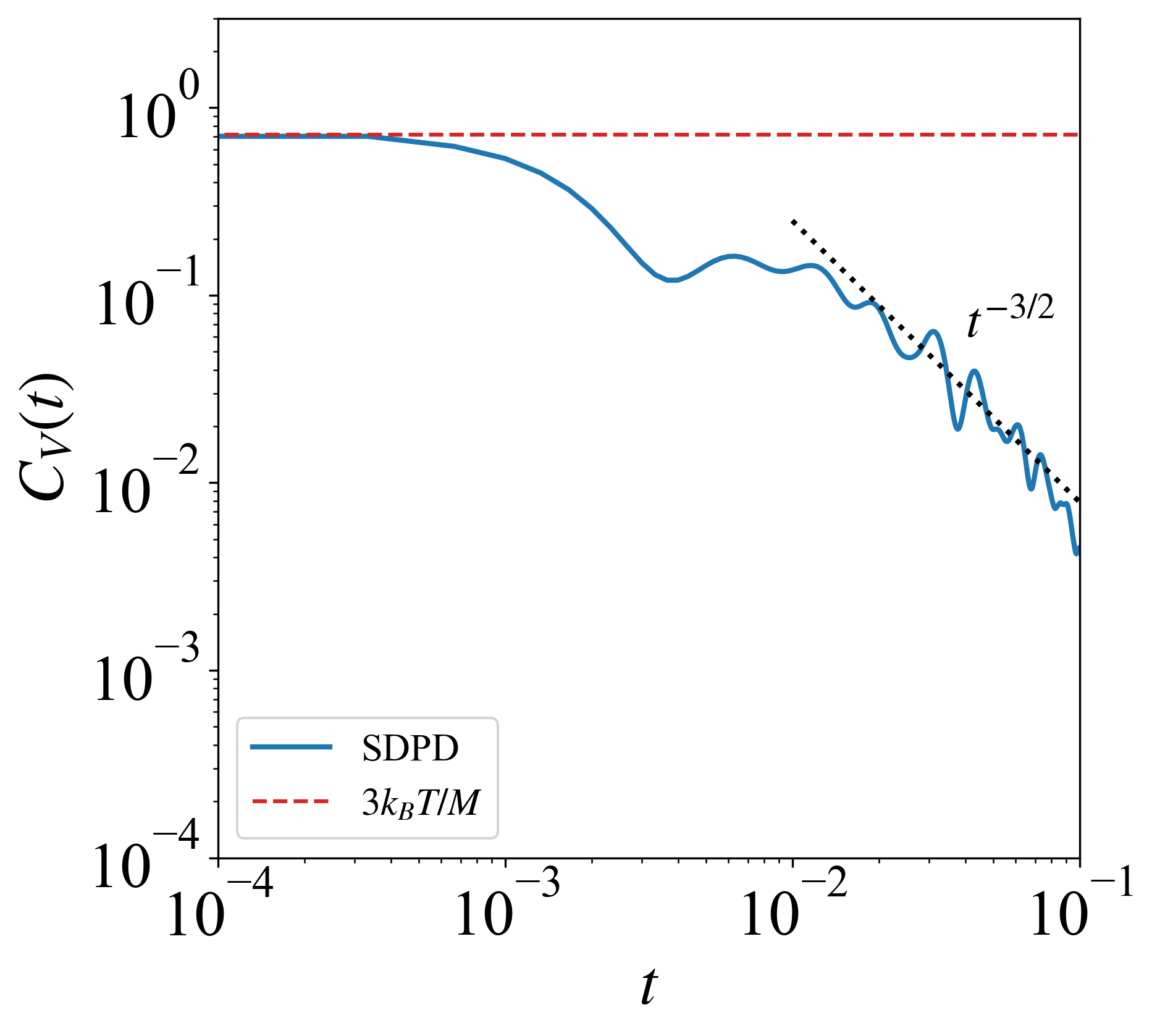}  
        \caption{VACF of a passive sphere ($B_n=0$).}  
        \label{fig:vacf}  
    \end{subfigure}
    \begin{subfigure}{0.45\textwidth}  
        \centering  
        \includegraphics[width=\textwidth]{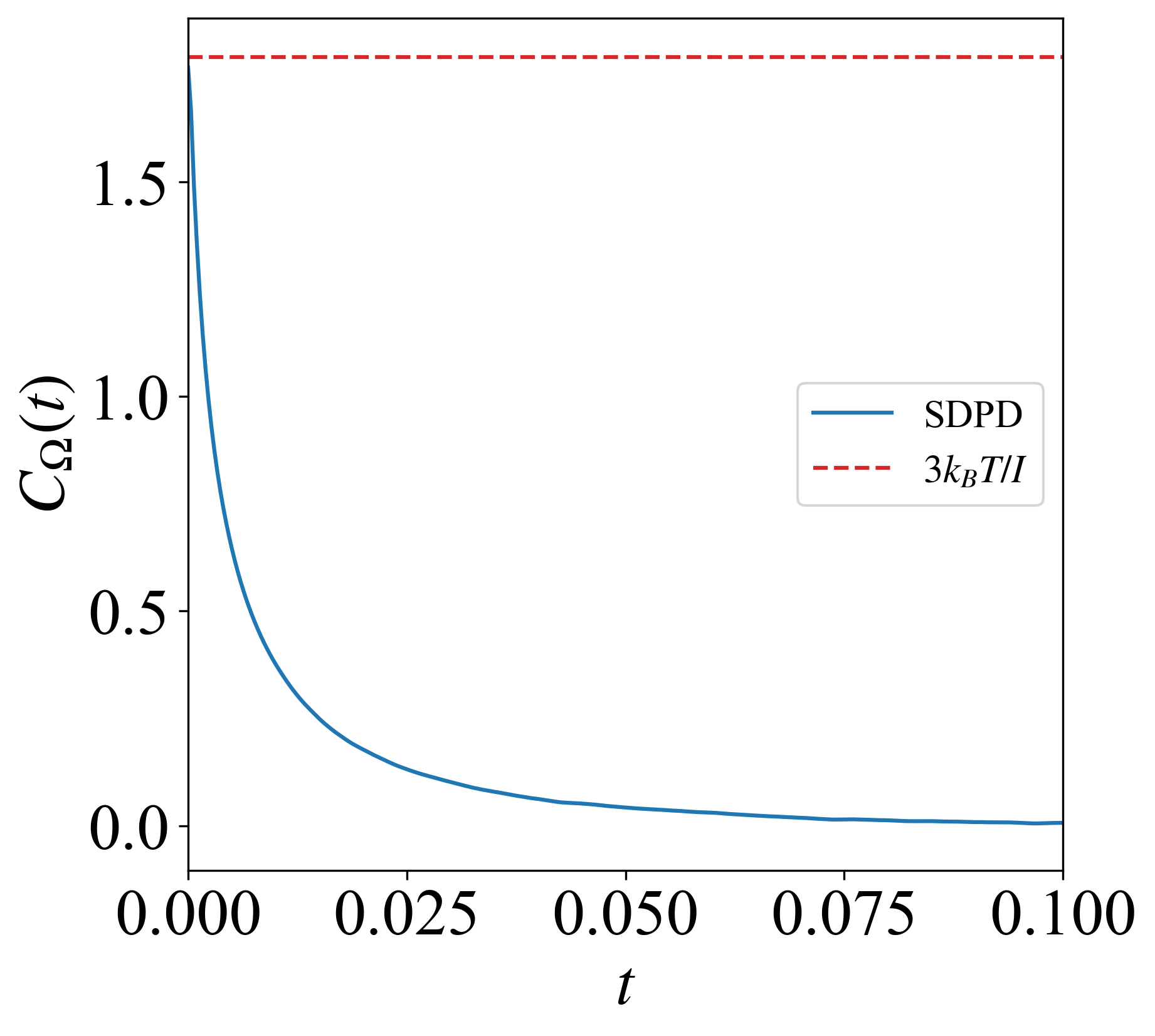} 
        \caption{AVACF of a squirmer ($B_1=3$).}  
        \label{fig:avacf}  
    \end{subfigure}  
    \caption{VACF of a passive sphere and AVACF of a squirmer.}  
    \label{fig:mesoscale}  
\end{figure}

Our initial validation involves assessing the velocity autocorrelation function (VACF) for a sphere that is not actively swimming ($B_1=0$), thus having a swimming velocity of zero. The VACF is mathematically expressed as $C_V(t) = \langle \mathbf{U}_0(t) \cdot \mathbf{U}_0(0) \rangle $, where the brackets denote an ensemble average. 
Fig.~\ref{fig:vacf} depicts the average result, which derived from 20 independent simulation runs. At the initial moment $t = 0$, the VACF aligns with the equipartition theorem from equilibrium statistical mechanics, which posits that $C_V(0) = \frac{3k_BT}{M} $, with $M$ representing the sphere's mass. As time progresses, the VACF exhibits a characteristic decay rate of $-3/2$.



We then test the effect of thermal fluctuations on the rotation of a single squirmer. Rotational diffusion plays a crucial role in the behaviour of the squirmer. The rotational P\'eclet number ($Pe_r$) is a dimensionless parameter that determines the dominance of rotational effects relative to advective effects. We define the rotational P\'eclet number of the squirmer as 
\begin{eqnarray}
    Pe_r = \frac{U_0}{2RD_r}
\end{eqnarray}
where $U_0$ is the unperturbed steady-state velocity of the squirmer and  $D_r$ is the rotational diffusion coefficient, defined as: 
\begin{eqnarray}
    D_r = \frac{k_BT}{8\pi \eta R^3}
\end{eqnarray}
In the case of the squirmer possessing an unperturbed steady-state velocity of $ U_0 = 2 $, the corresponding rotational P\'eclet number is calculated to be $ Pe_r = 25.1 $. For this scenario, the angular velocity autocorrelation function (AVACF) is measured, which is mathematically formulated as $ C_{\Omega}(t) = \langle \mathbf{\Omega}_0(t) \cdot \mathbf{\Omega}_0(0) \rangle $. The graphical representation of the average AVACF, derived from an ensemble of 20 independent simulations with different seeds, is depicted in Fig.~\ref{fig:avacf}. This result demonstrates that at the initial moment $ t = 0 $, the AVACF aligns with the principles of the equipartition theorem, specifically expressed as $ C_{\Omega}(0) = \frac{3k_BT}{I} $, with $ I $ denoting the moment of inertia of the sphere. This agreement signifies the fundamental relationship between the thermal energy and the rotational kinetic energy of the squirmer at the outset of its motion.

\subsection{Dynamics of  squirmer in multi-phase flow}

We extend the squirmer model to multi-phase flows, incorporating surface tension between different phases. In macroscopic or mesoscopic flows, the multiphase forces induced by surface tension may be comparable to, or even dominate over, inertial forces, thus necessitating the consideration of such forces. In this scenario, the squirmer navigates through a multi-phase fluid environment, where we assume the squirmer constitutes a distinct phase from the other two fluids, with thermal perturbation effects being neglected.

\begin{figure}[b]
\includegraphics[scale=0.5]{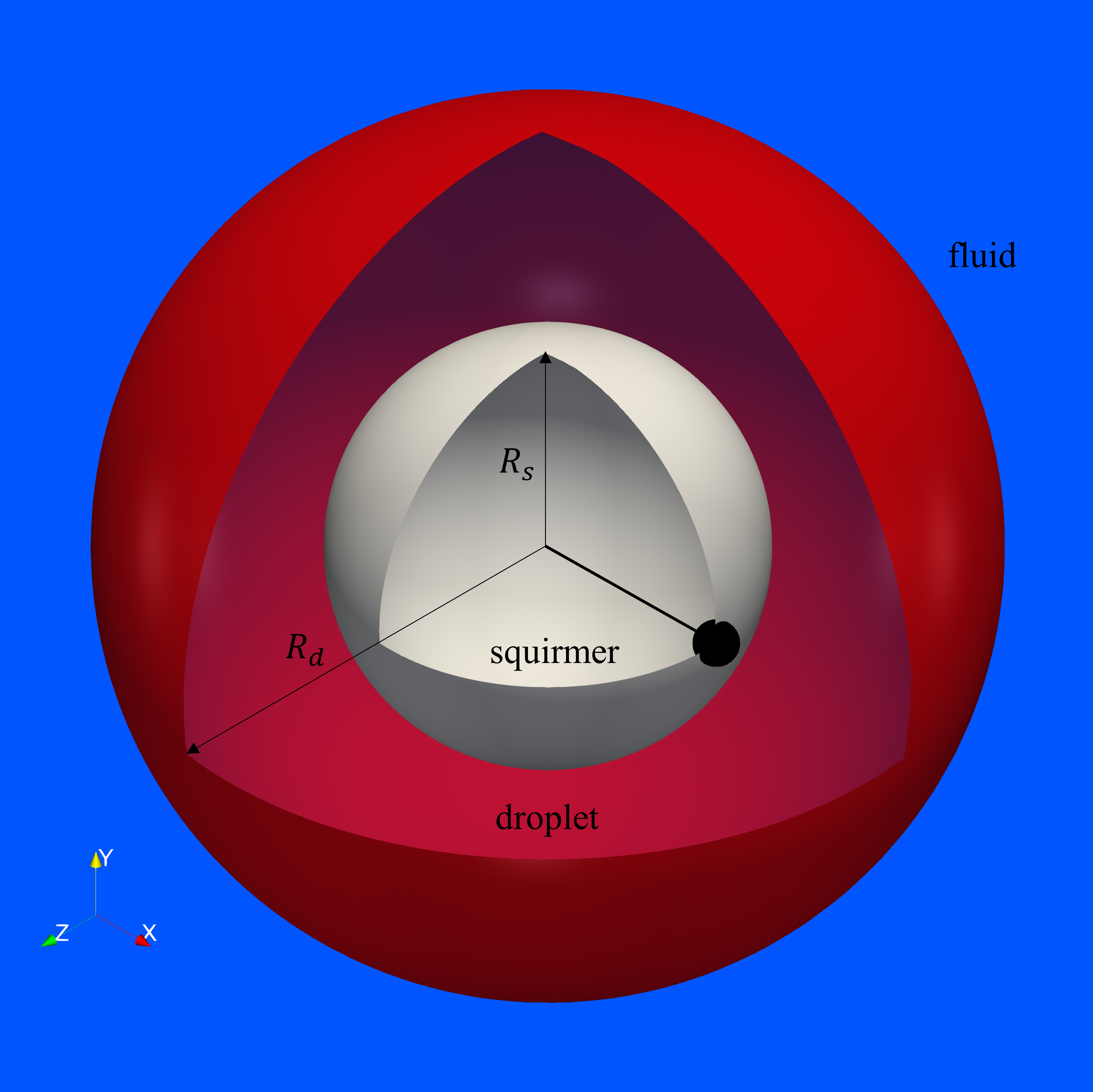}
\caption{\label{fig:mutiphase_confi}  Initial configuration in multiphase flow. A squirmer (white) with radius  $R_s$ is encapsulated by a droplet (red) with a radius $R_d$. This combination is immersed in a distinct fluid (blue).}
\end{figure}

Initially, the squirmer is encapsulated in a droplet, surrounded by another fluid. 
Fig.~\ref{fig:mutiphase_confi} illustrates the initial configuration in our three-dimensional simulation setup. For clarity, we denote the squirmer, droplet, and fluid with the letters $s$, $d$, and $f$, respectively. The radius of the squirmer is $R_s=1$. The swimming parameter is $B_1=0.75$ and the steady velocity magnitude in the bulk is $U_0=0.5$. The droplet is a sphere with a radius $R_d$. The computational region is a cubic box with a side length of $L = 20 R_s$ and all boundaries are periodic.
The densities of the droplet, fluid, and squirmer are identical, set to $\rho_d = \rho_f = \rho_s = 1$. The dynamics viscosities of the droplet and fluid are $\eta_d=0.25$ and $\eta_f=0.5$. The surface tension coefficients between the phases are $\alpha^{df} = 10$, $\alpha^{sf} = 10$, and $\alpha^{ds} = 5$. The corresponding Reynolds number for the squirmer moving in an undisturbed fluid is $Re=1$. We evaluated the squirmer for three different droplet sizes: large ($R_d = 3$), medium ($R_d = 2$), and small ($R_d = 1.5$). The Capillary number is $Ca = \frac{2\eta_f B_1}{3\alpha^{df}} = 0.025$.

\begin{figure*}[b]
 \includegraphics[scale=0.65]{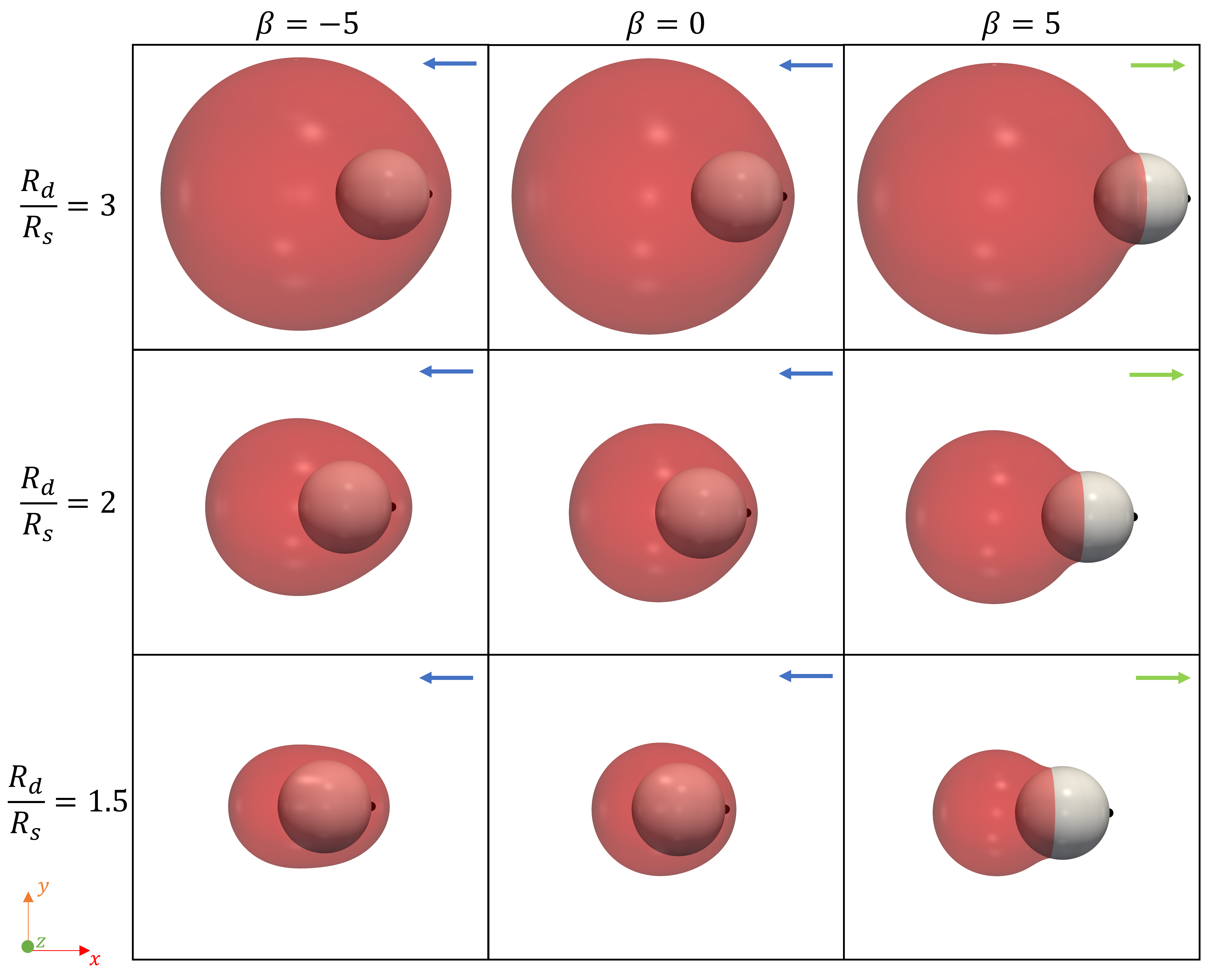}
\caption{\label{fig:mutiphase_types} Types of co-swimming motion of different squirmer with different droplet sizes at steady state. The white sphere represent squirmer, the red represent the droplets. The arrows indicate the direction of co-swimming of the droplet with the squirmer.}
\end{figure*}

\begin{figure*}[htbp]
    \centering  
    \begin{subfigure}{0.325\textwidth}  
        \centering  
        \includegraphics[width=\textwidth]{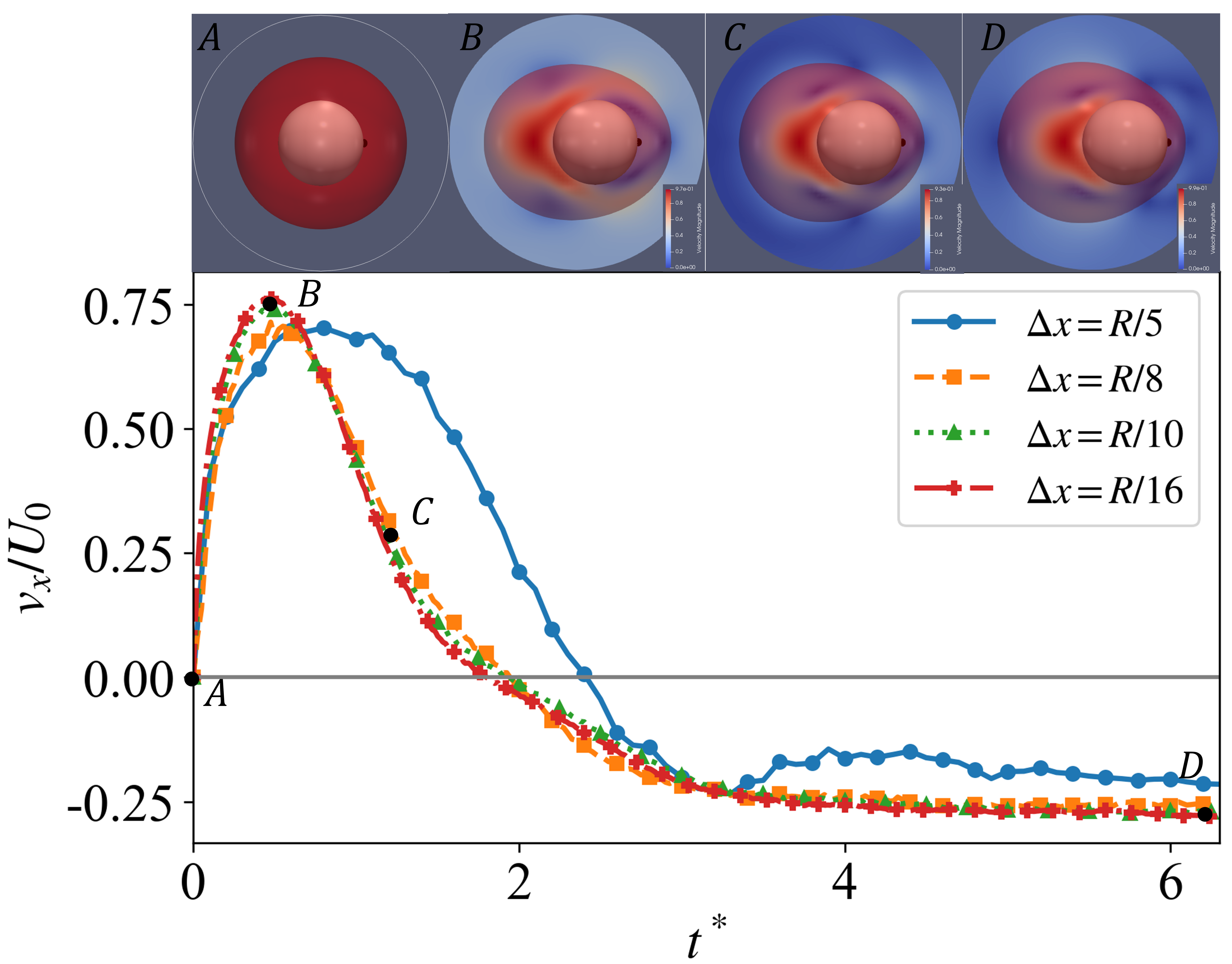}  
        \caption{$\beta=-5$}  
        \label{}  
    \end{subfigure}  
    \begin{subfigure}{0.315\textwidth}  
        \centering  
        \includegraphics[width=\textwidth]{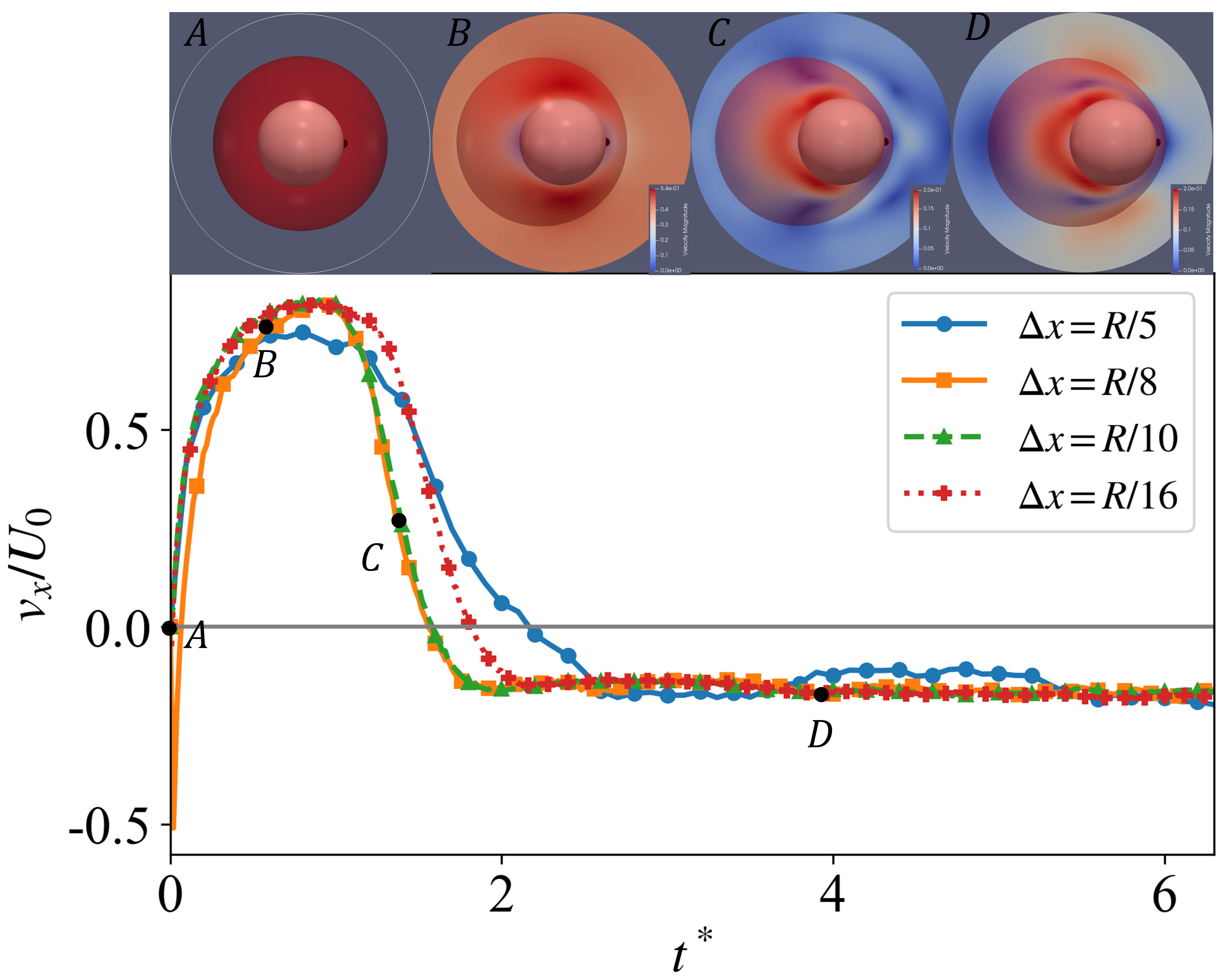}  
        \caption{$\beta=0$}  
        \label{}  
    \end{subfigure}  
    \begin{subfigure}{0.3\textwidth}  
        \centering  
        \includegraphics[width=\textwidth]{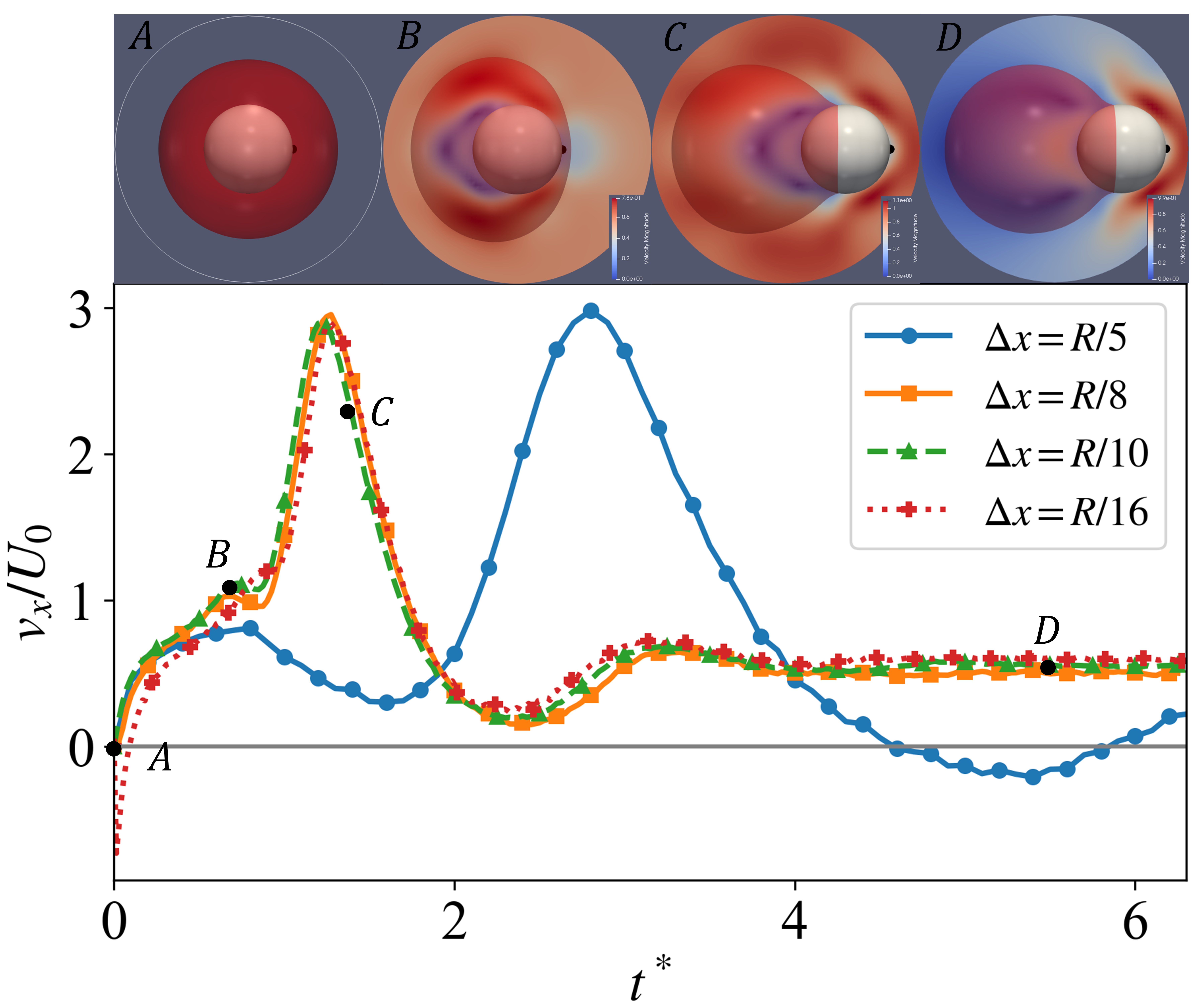}  
        \caption{$\beta=5$}  
        \label{}  
    \end{subfigure}  
    \caption{The evolution of the squirmer's velocity in the $x$ direction. The horizontal coordinate is the dimensionless time $t^*=tU_0/R$. The initial radius of droplets is $R_d=2$. The velocity field at several moments is displayed at the top.}  
    \label{fig:squirmer_droplet_velocity_time}  
\end{figure*}

Fig.~\ref{fig:mutiphase_types} shows the types of motion of squirmers with different droplet sizes at steady state. The orientation of the squirmer's head is consistently aligned with the positive x-axis direction. The droplet and squirmer eventually co-swim, which can be categorized into two distinct co-swimming behaviors. The first behavior is observed with pusher and neutral squirmers, which, irrespective of droplet size, co-swim in a direction opposite to the head's orientation.
Upon initiating motion from rest, the squirmer initially moves in the positive $x$-axis direction, aligning with the head's direction. As it approaches the droplet's edge, the squirmer slows down and reverses direction, eventually moving in the negative direction of the $x$-axis with the droplet. Throughout the motion, the squirmer never left the droplet. The second type of motion is produced by the puller, which eventually aligns with the head of the squirmer regardless of droplet size. Initially, the squirmer moves faster than the droplet, then it punctures through the droplet, and subsequently drags the droplet along the positive x-axis, with a part of its body still encapsulated within the droplet. Fig.~\ref{fig:squirmer_droplet_velocity_time} depicts the temporal variation of the squirmer's velocity in the $x$ direction for medium-sized droplets with $R_d=2$. The co-swimmer formed by the puller exhibits the greatest velocity magnitude at steady state, while the pusher demonstrates a slightly faster swimming speed than the neutral. The results converge as the resolution increases.

\begin{figure*}[b]
\includegraphics[scale=0.6]{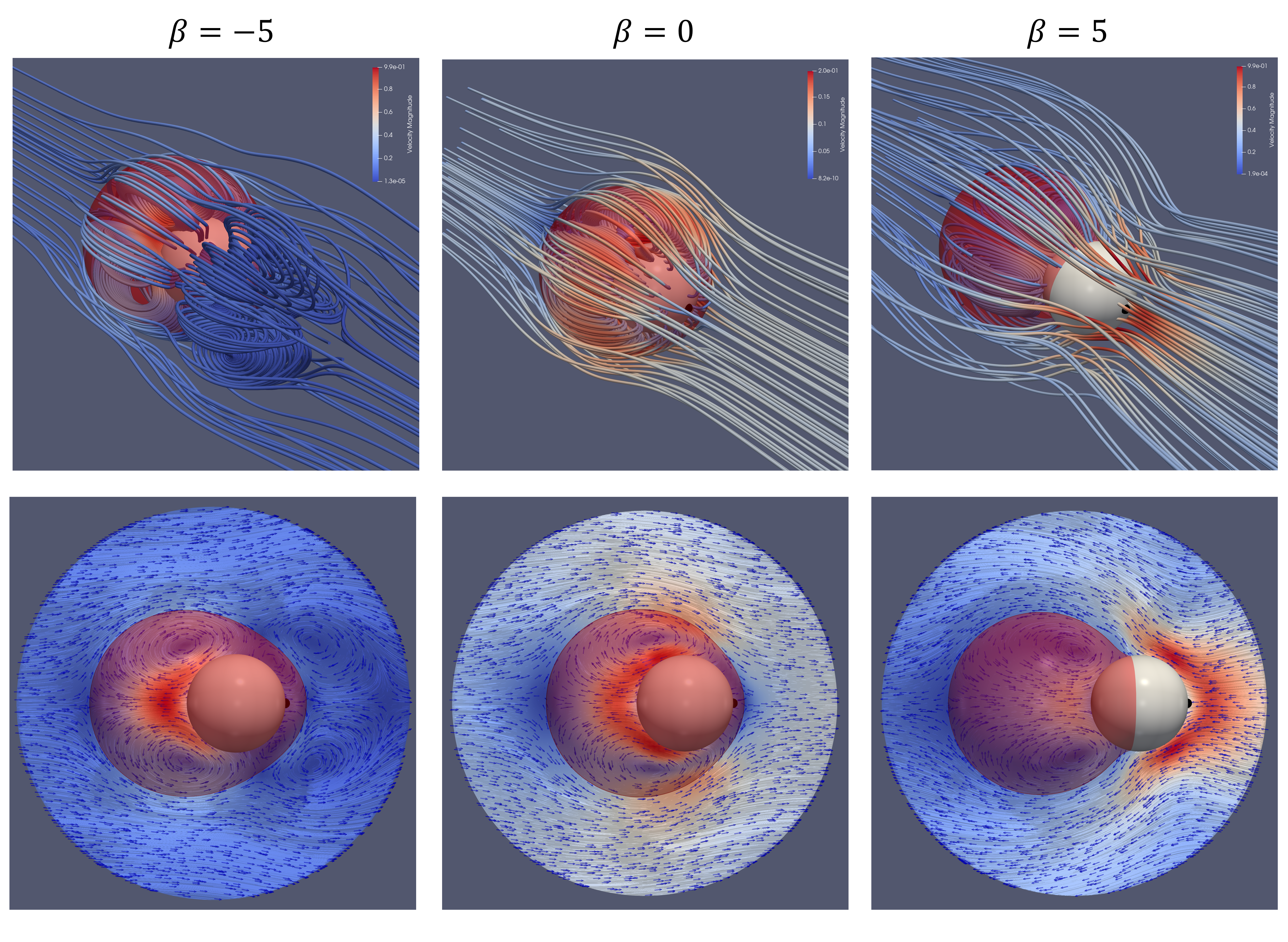}
\caption{\label{fig:multiphase_streamline}The stream lines and velocity field for a pusher ($\beta=-5$), a neutral squirmer ($\beta=0$) and a puller ($\beta=5$ ) in multi-phase flow in the frame moving with the swimmer.}
\end{figure*}

Fig.~\ref{fig:multiphase_streamline} illustrates the streamlines and velocity fields generated by squirmers and middle droplets at steady state. There is less variation for the puller compared to the flow field in the absence of droplets. For both the pusher and the neutral swimmer, the additional vortexes are generated at the tail of the squirmer, within the droplet. Treating the swimmer-droplet combination as a unified entity, the flow field structure suggests that the pusher-droplet combination mimics a puller oriented with its head in the negative $x$-axis direction, the neutral swimmer-droplet combination mimics a neutral squirmer also oriented negatively along the $x$-axis, and the puller-droplet combination mimics a neutral swimmer with its head facing the positive $x$-axis direction.

\section{Conclusion}\label{section_conclusion}
In this study, we apply the SPD method to model squirmer for the first time. The slip velocity is incorporated at the interface between the fluid and the spherical particle,
which is realized by assigning appropriate artificial velocities to the boundary particles of SPD. 
We accurately obtain the swimming velocity of a single squirmer at steady state and the surrounding flow field it generates. The resolution study shows convergence of our results and the flow field comply to the correct decay. Furthermore, we also simulate multi-squirmer collisions and the motion of a squirmer near a wall,
results of which are consistent with the literature. 
At the mesoscopic scale, where thermal perturbations are present, we obtain correct velocity and angular velocity autocorrelation functions. Finally, we extend the model to multiphase flows by considering a squirmer wrapped around by a droplet and meanwhile imposing a surface tension between the two flow phases. We find that the combination of a squirmer and a droplet with different physics properties exhibits distinct dynamic patterns. The proposed squirmer model has a potential to simulate a wide range of macroscopic and mesoscopic scenarios.

\begin{acknowledgments}
The authors acknowledge support from the National Natural Science Foundation of China under grant number: 12172330 and 12372264. Gaojin Li also acknowledges the support of the Natural Science
Foundation of Shanghai (grant number 23ZR1430800).
\end{acknowledgments}

\appendix

\section{Velocity field around a squirmer}\label{Velocity_field_analytic}


The velocity field around a squirmer of radius $R$ at a position $\mathbf{r}$ in Stokes flow is~\cite{blake1971_JFluidMech, ishikawa2006_et_al_JFluidMech}
\begin{widetext}
\begin{equation}
    \begin{aligned}
    \mathbf{v}&=-\frac{1}{3}\frac{R^{3}}{r^{3}}B_{1}\mathbf{e}+B_{1}\frac{R^{3}}{r^{3}}\frac{\mathbf{e}\cdot \mathbf{r}}{r}\frac{\mathbf{r}}{r}
    +\sum_{n=2}^{\infty}\left(\frac{R^{n+2}}{r^{n+2}}-\frac{R^{n}}{r^{n}}\right)B_{n}P_{n}\left(\frac{\mathbf{e}\cdot \mathbf{r}}{r}\right)\frac{\mathbf{r}}{r} \\
    &+\sum_{n=2}^{\infty}\left(\frac{n}{2}\frac{R^{n+2}}{r^{n+2}}-\left(\frac{n}{2}-1\right)\frac{R^{n}}{r^{n}}\right)B_{n}W_{n}\left(\frac{\mathbf{e}\cdot \mathbf{r}}r\right)\left(\frac{\mathbf{e}\cdot \mathbf{r}}r\frac{\mathbf{r}}{r}-\mathbf{e}\right). 
\end{aligned} 
\label{velocity_field_squirmer}
\end{equation}
\end{widetext}
where $r=|\mathbf{r}|$ is the distance from the center of the squirmer, $\mathbf{e}$ is the 
orientation of the squirmer. $P_n$ is the nth Legendre polynomial and $W_n$ is deﬁned by
\begin{equation}
    W_n(cos\theta) = \frac{2}{n(n+1)}P_n^{'}(cos\theta)
\end{equation}

\section{Box size study}\label{appendix_box_size}

\begin{figure*}[h]
    \centering  
    \begin{subfigure}{0.32\textwidth}  
        \centering  
        \includegraphics[width=\textwidth]{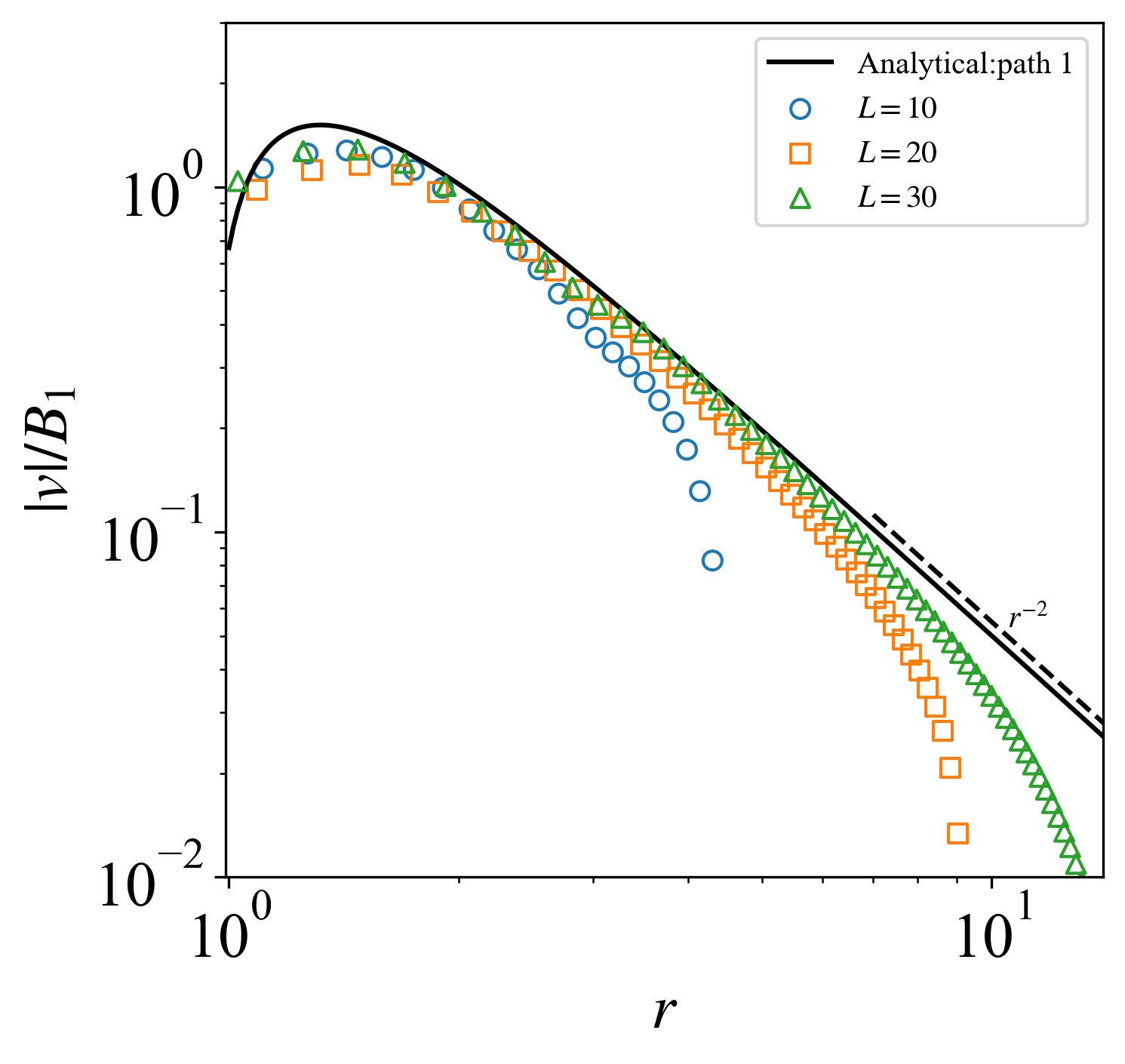}  
        \caption{Path 1}  
 
    \end{subfigure}  
    \begin{subfigure}{0.32\textwidth}  
        \centering  
        \includegraphics[width=\textwidth]{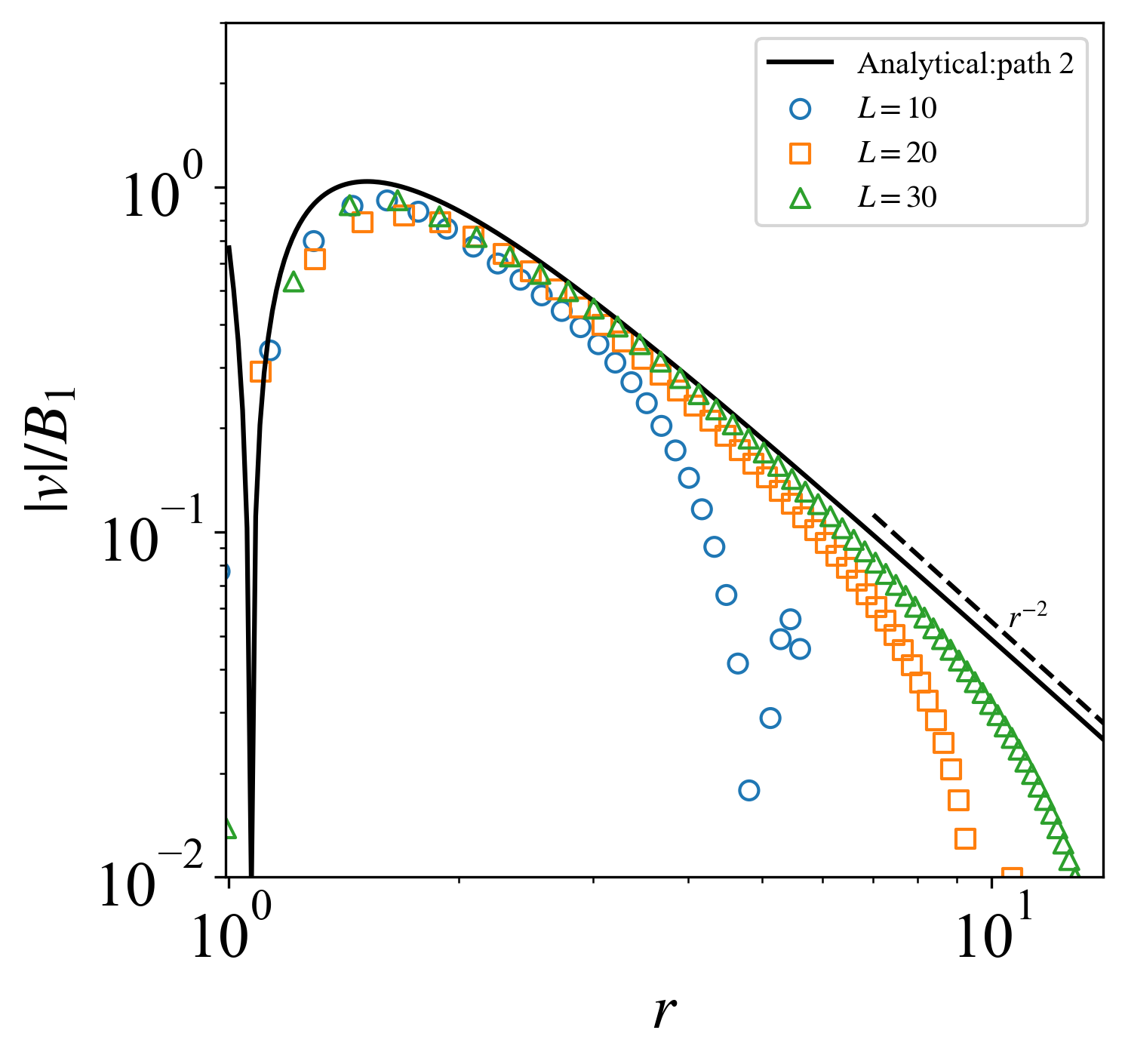}  
        \caption{Path 2}  
 
    \end{subfigure}  
    \begin{subfigure}{0.32\textwidth}  
        \centering  
        \includegraphics[width=\textwidth]{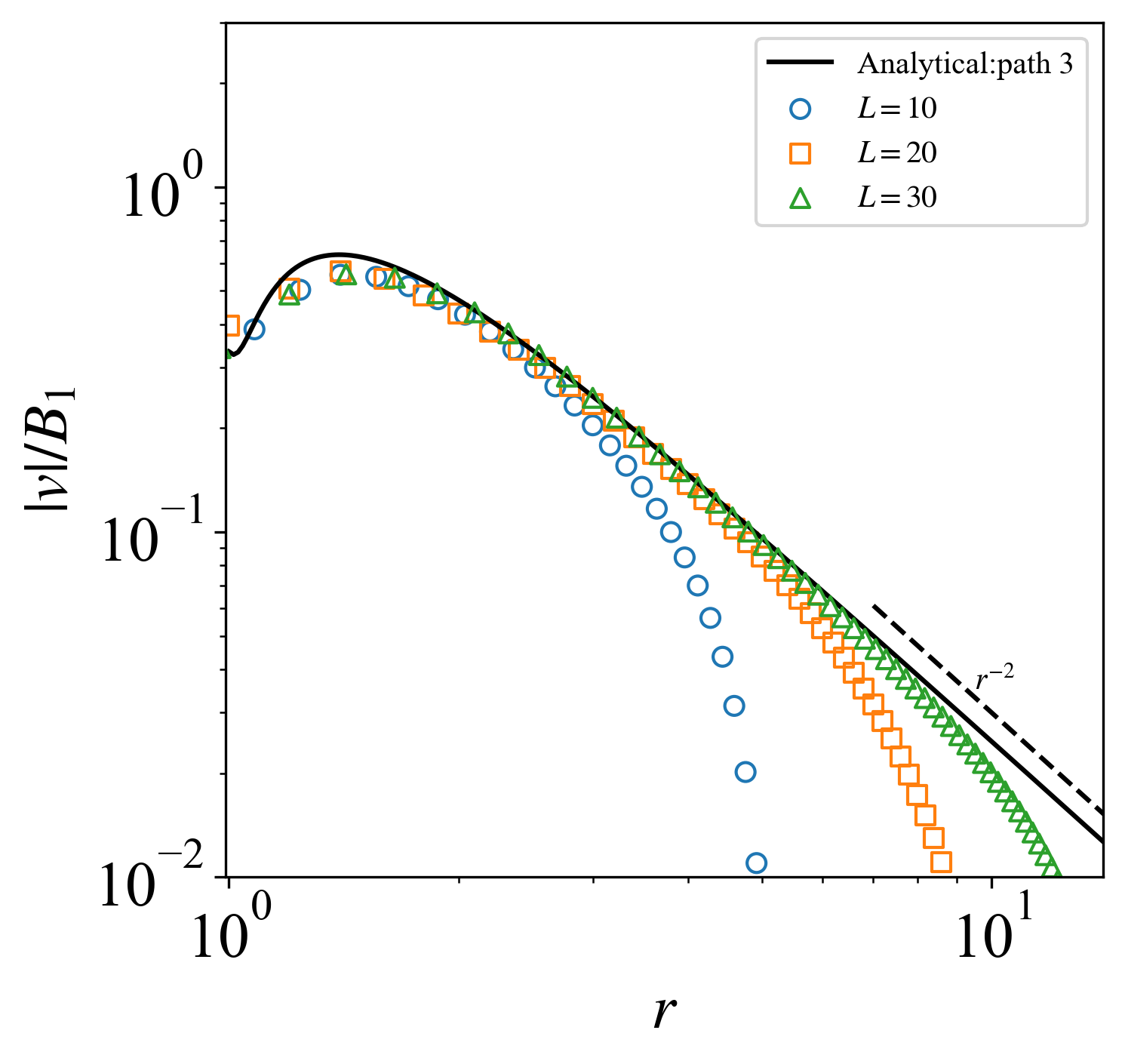}  
        \caption{Path 3}  
  
    \end{subfigure}  
    \caption{Box size study: The convergence of the flow field generated by an individual squirmer with increasing box size. (a) Path 1 is along the negative swimming direction, (b) path 2 is along the positive flow direction and (c) path 3 is perpendicular to the flow direction. The black solid lines represent the analytic solution of the velocity filed around a squirmer in bulk. The scatter represents the present simulation results for cubic boxes with different side lengths $L$.}  
    \label{fig:box_learning_decay_flow_field}  
\end{figure*}  

Periodic boundary conditions can influence the velocity decay of the flow field generated by an individual squirmer, causing an advance in the decay profile far away from the squirmer. In this study, we examine the impact of the simulation box size on the velocity decay characteristics. Fig.~\ref{fig:box_learning_decay_flow_field} presents the results for different sizes of cubic simulation boxes, with a constant simulation resolution of $R/dx = 10$ and a Reynolds number of $0.01$. As the box size increases, the results demonstrate convergence.


\bibliography{main}

\end{document}